\documentclass[12pt]{article}
\usepackage{epsf,amsfonts,amssymb,epsfig,amsmath,mathtools,graphics,slashed}
\usepackage{hyperref,hep,graphicx,subfig}
\usepackage{rotating}
\usepackage{xcolor}
\usepackage{verbatim}
\usepackage[final]{showlabels}

\newcommand{\nn}{\notag \\}

\newcommand{\ves}{\varepsilon}

\makeatletter

\@addtoreset{equation}{section}
\makeatother

\begin{document}

\begin{titlepage}

\vfill


\vfill

\begin{center}
   \baselineskip=16pt
   {\Large\bf Critical Dynamics of Superfluids}
  \vskip 1.5cm
  \vskip 1.5cm
      Aristomenis Donos and Polydoros Kailidis\\
   \vskip .6cm
   \begin{small}
      \textit{Centre for Particle Theory and Department of Mathematical Sciences,\\ Durham University, Durham, DH1 3LE, U.K.}
   \end{small}\\            
\end{center}

\vfill

\begin{center}
\textbf{Abstract}
\end{center}
\begin{quote}

We use standard techniques of hydrodynamics to construct a relativistic effective field theory for the low energy dynamics of nearly critical superfluids. In an appropriate non-relativistic limit, our theory predicts an additional coefficient when compared and contrasted to earlier work of Khalatnikov and Lebedev. In addition, we provide an alternative derivation of the same effective theory, using the Keldysh-Schwinger framework for non-equilibrium systems. Finally, we comment on the comparison with the results of an appropriate holographic computation presented in a companion paper. This provides further evidence in support of the theory we propose and confirms the existence of the extra coefficient we identified.
\end{quote}

\vfill

\end{titlepage}

\setcounter{equation}{0}

\section{Introduction}

Hydrodynamics provides a universal framework for studying finite temperature many-body systems, out of thermodynamic equilibrium \cite{Landau1987Fluid,Kovtun:2012rj,Romatschke:2017ejr}. At sufficiently long time and length scales\footnote{Compared to typical microscopic scales, such as the mean free path and mean free time.} all microscopic degrees of freedom have equilibrated, leaving conserved charges to dominate the effective description. The intuitive reason behind this is that conserved charges, by definition, cannot be destroyed locally and need to spread out across the system, in order to equilibrate.

The situation is drastically changed in the vicinity of a second order phase transition. In this case, the order parameter driving the transition is an additional slow variable due to the phenomenon of critical slowing down \cite{Landau:1954han}. A manifestation of the breakdown of conventional hydrodynamics in this case is that various transport coefficients diverge close to the critical point. A consistent description of the critical dynamics requires the inclusion of the order parameter in the variables of the effective description \cite{Stephanov:2017ghc, STEPHANOV2017876}.

The focus of this paper will be on the nearly critical dynamics of superfluids, with the order parameter being a complex scalar field, charged under a global $U(1)$ symmetry. The goal is to construct a set of equations that couple the order parameter to the standard hydrodynamic degrees of freedom and describe their time evolution, including the effects of dissipation. The first attempt to write down such an effective theory was made a long time ago by Pitaevskii \cite{Pitaevskii1959}. Subsequent related works include \cite{Khalatnikov1970, PhysRevB.22.3207, KhalatnikovLebedev1978}. (For a general review, see \cite{GinzburgSobyanin1982}.) The most general effective description was given by Khalatnikov and Lebedev in \cite{KhalatnikovLebedev1978}, in the non-relativistic case. However, one of the main results of our present work is showing that the order parameter equation put forward in \cite{KhalatnikovLebedev1978} lacks a specific term. This term contributes at next-to-leading order in a certain perturbative expansion, which we outline in the main text. As we discuss in the main text, this term is necessary in the effective description of holographic superfluids. The holographic computation is presented in \cite{Donos:2025igh}.

It is important to emphasise that any theory of critical dynamics should not only be concerned with the description of the system at the level of classical equations of motion. Close to the critical point, thermal fluctuations are amplified and a renormalisation group treatment is necessary for both static and dynamical phenomena. An important milestone in this respect has been the review by Hohenberg and Halperin \cite{RevModPhys.49.435}, that classified the dynamics of various nearly critical systems and discussed their RG analysis. (See also \cite{Folk_2006} for a more recent review and \cite{osti_7362153, Vasilev:2004yr} for a pedagogical approach.)

On a similar note, the classical equations of motion of hydrodynamics provide only a first approximation for the local dynamics of conserved charges and Goldstone modes. It is well known that the correlators of conserved currents seen in low-energy experiments and simulations have power-law long-time tails \cite{PhysRevA.1.18,POMEAU197563}, contradicting the prediction of exponential decay due to hydrodynamic dissipation \cite{Romatschke:2017ejr}. The power-law behaviour is a hallmark of fluctuations, which can be systematically described in the complete framework of the Keldysh-Schwinger formalism \cite{Crossley:2015evo, Haehl:2015uoc, Liu:2018kfw}. Classical hydrodynamics can be then viewed as a saddle point approximation.

With these considerations in mind, we have also constructed a Keldysh-Schwinger effective action for our system, as a first step in including the effect of fluctuations. Despite this, our main interest will be centered on the mean field level equations governing our nearly critical system.

Holography \cite{Aharony:1999ti,Witten:1998qj, Hartnoll:2016apf } is another well-established framework for studying the hydrodynamic limit of various systems from first principles \cite{Policastro:2002se, Baier:2007ix, Bhattacharyya:2008jc}. In order to put our effective theory to the test, in a separate work \cite{Donos:2025igh} we have performed an analytic holographic computation to extract the effective theory of nearly critical holographic superfluids. As we explain in the main text, our findings there validate the theory proposed in this work, in an appropriate mean field theory limit.

This paper is organised as follows. In Section \ref{sec:Eff_th} we construct the effective theory of nearly critical superfluids using a hydrodynamic approach and proceed to linearise it around a homogeneous background without superfluid velocity. In Section \ref{sec:Asymptotic} we consider two asymptotic limits of the linearised theory, namely for small and large values of the wavevector modulus compared to the gap of the amplitude mode. We show explicitly that the effective theories of superfluids and charged normal fluids are obtained in these limits. In Section \ref{sec:KS_hydro} we present an alternative, Keldysh-Schwinger construction of the same effective theory, including thermal fluctuations of the effective degrees of freedom. We conclude this work with a discussion in Section \ref{sec:Disc}. 

\section{Effective theory near the critical point}\label{sec:Eff_th}

In this section we construct an effective theory for superfluids, valid arbitrarily close to the critical point of the second order phase transition. The variables of the theory are going to include the usual set of hydrodynamic variables of a normal fluid, the temperature $T$, the chemical potential $\mu$ and the normal fluid velocity $u^\mu$  (normalized as $u^\mu u_\mu=-1$). In addition, we will need a complex scalar field $\psi$, charged under a $U(1)$ symmetry, playing the role of the order parameter that drives the superfluid transition. The superfluid phase exists below a certain critical temperature $T_c$, in which the field $\psi$ acquires a non-zero expectation value, breaking the $U(1)$ symmetry of the theory spontaneously.

 Our goal is to describe the low-energy dynamics of the effective degrees of freedom, in the background of a $d$-dimensional spacetime with fixed metric $g_{\mu\nu}$, an external $U(1)$ gauge field $A_\mu$, and an external complex source $s_\psi$, thermodynamically conjugate to the order parameter $\psi$. All the external sources are taken to be slowly varying functions of the spacetime coordinates. In general terms, we can introduce the partition function of the system, $Z[g_{\mu\nu},A_\mu,s_\psi]=e^{ W[g_{\mu\nu},A_\mu,s_\psi]}$, with $W$ being the generating functional.  The stress tensor and $U(1)$ current of the theory can then be obtained by taking functional derivatives as,
\begin{align}\label{eq:func_der}
     T^{\mu\nu}=\frac{2}{\sqrt{-g}}\frac{\delta W}{\delta g_{\mu\nu}}\,,\quad J^\mu=\frac{1}{\sqrt{-g}}\frac{\delta W}{\delta A_\mu}\,.
 \end{align}

Given these definitions and assuming the absence of anomalies, the invariance of $W$ under diffeomorphisms and $U(1)$ gauge transformations leads to the continuity equations,
\begin{align}\label{eq:contin}
   \nabla_\mu T^{\mu\nu}=&F^{\nu\mu}J_\mu+\frac{1}{2}\left(\psi^\ast\,D^\nu s_\psi +\psi\,D^\nu s_\psi^\ast\right)\,,\nn
   \nabla_\mu J^\mu=&\frac{q_e}{2i}\left(\psi^\ast\,s_\psi-s_\psi^\ast\,\psi\right)\,.
\end{align}
In the above equations, we have introduced the $U(1)$ covariant derivative, $D_\mu\psi=\nabla_\mu\psi+iq_eA_\mu\psi$ and similarly for $s_\psi$. In the broken phase, one can also introduce the polar decompositions,
\begin{align}
   \psi=|\psi|e^{i q_e \theta},\quad s_\psi=|s_\psi|e^{i q_e \theta_s} \,.
\end{align}
 An effective theory for the system we wish to describe must provide us with the constitutive relations, which express the stress tensor and current in terms of the $d+3$ independent hydrodynamic variables $\mu, T,u^\mu,\psi$, and the external sources $g_{\mu\nu},\,A_\mu,\,s_\psi$. Along with the $d+1$ continuity  equations \eqref{eq:contin}, we will need two more to close the system of equations. These two can be jointly thought of as a single complex equation that captures the dynamics of the order parameter $\psi$\footnote{This is similar to the case of conventional superfluids, where along with the continuity equations, we have the Josephson relation, describing the dynamics of the phase field \cite{Bhattacharya:2011tra}.}.

\subsection{Equilibrium}

As a first step, in this subsection it will be important to understand the state of thermodynamic equilibrium in our system, in a local sense. Similar techniques have been applied in the past in \cite{Banerjee:2012iz,Jensen:2012jh,Bhattacharyya:2012xi,Jensen:2013kka} 
to describe equilibrium (in the presence of external sources) in various systems. 

By definition, at equilibrium we can assume the existence of a timelike Killing vector $K^\mu$ and a $U(1)$ gauge transformation $\lambda_K$ such that the Lie derivative $\mathcal{L}_K$ on the external sources along $K$ satisfies\footnote{Notice that then the action on the gauge invariant combination of the external phase and gauge field is $\mathcal{L}_K(\nabla_\mu\theta_s+A_\mu)=0\,.$},
\begin{align}
    \label{eq:Lie_cond}
&\mathcal{L}_K\,g_{\mu\nu}=0,\quad\mathcal{L}_K A_\mu=\nabla_\mu\lambda_K,\quad\mathcal{L}_K|s_\psi|=0,\quad\mathcal{L}_K\theta_s=-\lambda_K,
\end{align}
for some well defined gauge transformation parameter $\lambda_K$ on our manifold. These conditions impose that a transformation along the flow of the Killing vector $K$ can induce a gauge transformation, at most. The equilibrium temperature, normal fluid velocity and chemical potential are then defined through,
\begin{align}
\label{eq:equil_vars}
    T=\frac{T_0}{\sqrt{-K^2}},\quad u^\mu=\frac{K^\mu}{\sqrt{-K^2}}, \quad \mu=\frac{K^\mu A_\mu-\lambda_K}{\sqrt{-K^2}},
\end{align}
with $T$ being the position dependent inverse length of the Euclidean thermal circle and $T_0$ the time coordinate periodicity. It is straightforward to check that the quantities  $T,u^\mu,\mu$ have zero Lie derivative with respect to $K$ and are therefore suitable variables to describe the system at equilibrium. In the same spirit, we impose the conditions,
\begin{align}\label{eq:order_par_Lie_cond}
    \mathcal{L}_K |\psi|=0,\quad\mathcal{L}_K\,\theta=-\lambda_K\,,
\end{align}
on the modulus and phase of our complex order parameter. It is reassuring to observe that the usual Josephson relation,
\begin{align}
    \mu = u^\mu (\nabla_\mu\theta+A_\mu)\,,
\end{align}
is a matter of combining equations \eqref{eq:equil_vars} and \eqref{eq:order_par_Lie_cond}. As a final comment, we will later see that it will be useful to introduce the modified covariant derivative $\hat{D}_\mu=\nabla_\mu+i q_e A_\mu +iq_e \mu\, u_\mu $, bringing the conditions \eqref{eq:order_par_Lie_cond} to the simple form,
\begin{align}\label{eq:Jos_like}
    u^\mu \hat{D}_\mu\psi=0\,.
\end{align}

We will now move on to the generating functional $W$. At equilibrium, we can imagine that $W$ is obtained from an associated Euclidean effective action $F$, after integrating out the condensate. Schematically,
\begin{align}\label{eq:W_path}
    e^W=\int d\psi\,d\psi^\ast e^{-F}\,.
\end{align}
The phase of the condensate, $\theta$, and its amplitude  $|\psi|$,  correspond to a gapless and nearly gapless degree of freedom, respectively. This suggests that close to the critical point, $W$ is going to be a non-local functional of the external sources. It will therefore  be more convenient to focus on $F$ instead\footnote{For a similar discussion in the case of superfluids away from the critical point, see \cite{Bhattacharyya:2012xi}.}. This is in contrast to normal fluids, for which in a static equilibrium $W$ is a local functional of the sources\cite{Jensen:2012jh, Banerjee:2012iz}. We further note that in this section we are primarily interested in the mean field theory limit, ignoring the all-important thermal fluctuations close to $T_c$. This in turn means that the path integral in \eqref{eq:W_path} is evaluated through the saddle-point approximation with the saddles satisfying,
\begin{align}\label{eq:saddle}
    \mathcal{F}_\psi\equiv\frac{1}{\sqrt{-g}}\frac{\delta F}{\delta \psi^\ast}=0\,,\quad \mathcal{F}^\ast_\psi\equiv \frac{1}{\sqrt{-g}}\frac{\delta F}{\delta \psi}=0\,.
\end{align}

As usual, we define the projection operator transverse to the normal fluid velocity as $ P^{\mu\nu}=g^{\mu\nu}+u^\mu u^\nu$\footnote{If the spacetime is static (not only stationary, as we have assumed) then in appropriate coordinates the velocity $u^\mu=\left(\frac{1}{\sqrt{-g_{00}}},\vec{0}\right)$ is the unit normal vector to the spacelike hypersurfaces $\Sigma_t:t=const.$ Also, in that case $P^{\mu\nu}$ projects an arbitrary spacetime tensor to a tensor tangent to $\Sigma_t$.}. We will also decompose the spacetime covariant derivative,
\begin{align}
    \nabla_\mu = \nabla_\mu^\perp - u_\mu\,\partial_u,
\end{align}
in the comoving time covariant derivative $\partial_u \equiv u^\mu \nabla_\mu$ and the transverse space derivative $\nabla_\mu^\perp\equiv P_\mu{}^\nu\,\nabla_\nu$. Similarly, we also identify $u^\mu \hat{D}_\mu\equiv\hat{D}_u$. It is reasonable to assume that the Euclidean effective action $F$ will be a functional of the transverse derivatives of our fields.

To achieve this, we write the effective action as an integral over a local density according to\footnote{In coordinates where $K^\mu=(1,\vec{0})$ we have  $F=\frac{1}{T_0}\int d^{d-1}x\sqrt{-g}\,f_{tot}$ since the integrand is independent of the time coordinate $x^0$.},
\begin{align}
    F=\int d^dx \sqrt{-g}\,f_{tot}\,.
\end{align}
At leading order in derivatives of $\psi$, we can express,
\begin{align}\label{eq:f_tot_def}
    f_{tot}=\frac{w_0(\mu,T)}{2}|D_\mu^\perp\psi|^2+f(\mu,T,|\psi|^2)-\frac{1}{2}\left(s_\psi^\ast\,\psi+s_\psi\,\psi^\ast\right)\,.
\end{align}
 At this order, the functional derivative in the equation of motion of the order parameter \eqref{eq:saddle} can be written as\footnote{The functional derivative with respect to $\psi$ is calculated keeping $g_{\mu\nu},A_\mu,s_\psi,T_0$ (and hence $K^\mu,\lambda_K$) fixed. Also, we only consider variations of $\psi$ that satisfy the Josephson relation, i.e.  $u^\mu D_\mu\,\delta\psi=0$.},
\begin{align}\label{eq:F_wrt_psi}
   \mathcal{F}_\psi^\ast=-T\,D_\mu^{\perp}\left(\frac{w_0}{2\, T}D^{\perp\mu}\psi^\ast\right)+\frac{\partial f}{\partial |\psi|^2}\psi^\ast-\frac{s_\psi^\ast}{2}\,.
\end{align}
To obtain this expression, we have used that in equilibrium $\nabla_\mu^{\perp}T=-T\,u^\nu\nabla_\nu u_\mu$, which follows from the definitions \eqref{eq:equil_vars} and the Killing equation for $K^\mu$.

In equilibrium, $f_{tot}$ can be interpreted as the local free energy in the local rest frame of the normal fluid component, while $F$ is nothing but the Ginzburg-Landau potential \cite{Ginzburg:1950sr, Ginzburg:1958}. In addition, we note that the function $f$ can be further expanded in powers of $|\psi|^2$, as discussed in the appendix; however, its explicit form will not be important in the following.

We can now apply \eqref{eq:func_der} and the formulas \eqref{eq:equil_vars}, assuming $\psi$ is on-shell, to get the constitutive relations for the current and stress tensor in equilibrium\footnote{Under variations of the metric and gauge field, while keeping $T_0,\, K^\mu,\,\lambda_K$ fixed, the definitions \eqref{eq:equil_vars} give, $\delta T=\frac{T}{2}u^\mu u^\nu \delta g_{\mu\nu}$, $\delta u^\mu=\frac{u^\mu}{2}u^\rho u^\sigma \delta g_{\rho\sigma}$, $\delta \mu=\frac{\mu}{2}u^\mu u^\nu \delta g_{\mu\nu}+u^\mu \delta A_\mu$.},
\begin{align}\label{eq:con_rel_therm}
    J^\mu_{eq}&=\varrho\, u^\mu+q_e w_0\, \mathrm{Im}(\psi D^{\perp\mu}\psi^\ast)\,,\nn
    T^{\mu\nu}_{eq}&=\epsilon\, u^\mu u^\nu+p\,P^{\mu\nu}+2\,w_0\,q_e\,\mu\,u^{(\mu}\,\mathrm{Im}\left(\psi D^{\perp\nu) }\psi^\ast\right)+w_0 D^{\perp(\mu}\psi D^{\perp\nu)}\psi^\ast\,,
\end{align}
where we have used the thermodynamic relations,
\begin{align}\label{eq:therm_def}
    \varrho=-\frac{\partial f_{tot}}{\partial \mu},\quad s=-\frac{\partial f_{tot}}{\partial T},\quad p=-f_{tot}, \quad \epsilon+p=s\,T+\mu\,\varrho\,,
\end{align}
with $\varrho,\,s,\,\epsilon,\,p$ being, respectively, the local charge, entropy, energy and pressure, in the local rest frame of the fluid at equilibrium.
As we have already stressed, $J^\mu_{eq}$ and $ T^{\mu\nu}_{eq}$ automatically satisfy the continuity equations \eqref{eq:contin}, if $\psi$ is on-shell, due to the coordinate and gauge invariance of the effective action. It is also worth noting that equation \eqref{eq:con_rel_therm} shows that the current-current susceptibility is given by,
\begin{align}\label{eq:cur_cur_susc}
    \chi_{JJ} = w_0\,q_e^2\,\left|\psi \right|^2\,.
\end{align}

The final ingredient we would like to discuss at equilibrium is the entropy current,
\begin{align}\label{eq:entr_can}
    s_{eq}^\mu=\frac{p}{T}u^\mu-T_{eq}^{\mu\nu}\frac{u_\nu}{T}-\frac{\mu}{T}J_{eq}^\mu=s\,u^\mu\,.
\end{align}
One can check that on-shell $\nabla_\mu s_{eq}^\mu=0$, so indeed there is no entropy production and the entropy current is carried by the normal fluid only, since $s_{eq}^\mu$ is proportional to $u^\mu$.

\subsection{Dissipation}\label{sec:Dissipation}

We shall now move on to take into account the effects of dissipation. Away from thermodynamic equilibrium, the stress tensor, the electric current and the equation of motion for the order parameter will receive dissipative corrections according to,
\begin{align}\label{eq:diss_mod}
    T^{\mu\nu}=T^{\mu\nu}_{eq}+T^{\mu\nu}_{diss},\quad J^\mu=J^\mu_{eq}+J^\mu_{diss},\quad  \hat{D}_u \psi=E_{diss}\,.
\end{align}
The main goal of this section is to write the leading contributions to $T^{\mu\nu}_{diss},\,J^\mu_{diss},\,E_{diss}$ in a consistent expansion scheme.

Away from equilibrium, the constitutive relations cannot be obtained from a variational principle. To do so, one must first include additional fields in the path integral computation of $W$, according to the Keldysh-Schwinger formalism for effective theories \cite{Liu:2018kfw}. Alternatively, in the spirit of conventional hydrodynamics, one writes down all possible corrections in the dissipative terms ($T^{\mu\nu}_{diss}$ etc.), which respect the diffeomorphism and $U(1)$ gauge invariance of the theory, and demand positivity of entropy production along with Onsager reciprocity\cite{Kovtun:2012rj}. This will be our approach in the rest of this subsection. Later, in Section \ref{sec:KS_hydro}, we also present a Keldysh-Schwinger construction of the theory.

The main technical difference compared to hydrodynamics will be due to the presence of the small quantity $\mathcal{F}_\psi$ in our system. Conventional hydrodynamics describe the dynamics of systems close to global thermodynamic equilibrium, i.e. close to a state in which all the hydrodynamic variables are constant in spacetime. In this limit, the gradients of the hydrodynamic variables quantify the departure from global equilibrium and naturally serve as the small expansion parameter of hydrodynamics. In our case, there is another quantity that is zero in equilibrium and therefore can also serve as a small expansion parameter when we are close to equilibrium, and this is $\mathcal{F}_\psi$. The conclusion is then that we must expand the constitutive relations of equation \eqref{eq:diss_mod} in terms that include derivatives of $\mu,T,u^\mu,\psi$ (and of the external sources) and appropriate factors of $\mathcal{F}_\psi$ and its derivatives.

Such a term, i.e. the derivative of some thermodynamic potential $F$ with respect to a non-hydrodynamic\footnote{Non-hydrodynamic in the sense that it is not associated with a conserved quantity.} mode $\chi$ was first used in \cite{mandelshtam1937theory} to write a relaxation equation for the non-hydrodynamic mode, of the form $\partial_t\chi\sim\frac{\partial F}{\partial \chi}$. (See also the discussion in Section 101 of  \cite{LIFSHITZ1981427}.) Such terms are also the building blocks of all the models of critical dynamics, as reviewed in \cite{RevModPhys.49.435}. Moreover, let us observe that in the framework of Keldysh-Schwinger effective theories, whenever one decides to include a non-hydrodynamic mode $\chi$ in the theory, a term $\frac{\partial F}{\partial \chi}$ appears naturally in the equation of motion of this mode, after imposing the so-called ``dynamical KMS symmetry''(see e.g. Appendix D of\cite{Glorioso:2016gsa} for model A and \cite{Donos:2023ibv} for the case of superfluids.) We will see this in detail in Section \ref{sec:KS_hydro}.

Our first task is the development of an expansion scheme for the possible dissipative terms entering \eqref{eq:diss_mod}. Conventional hydrodynamics describes the (slowest possible) relaxation of a finite-temperature system back to equilibrium. The time scale of such a process $\tau_{hydro}$ grows arbitrarily large as the spatial inhomogeneity of the fluctuations related to the conserved charges, denoted by $l_{inh.}$, grows arbitrarily large\footnote{ Equivalently, in Fourier space, the  corresponding hydrodynamic modes have frequency $\omega_{hydro}\sim\frac{1}{\tau_{hydro}}$ that vanishes as the wavevector $k=\frac{1}{l_{inh.}}\to 0.$}. This is an immediate corollary of the continuity equations, since e.g. from  $\partial_t T^{tt}=-\partial_i T^{ti}$, for small $\partial_i$, the time variation of the conserved charge $T^{00}$ becomes small as well.   All the rest of the (non-hydrodynamic) processes need time scales $\tau_{rest}\ll\tau_{hydro}$ to equilibrate the system.  This clear separation of scales is the reason why hydrodynamics universally captures the effective dynamics of a generic system at macroscopic time and length scales.

Close to a phase transition this picture is enriched due to the critical slowdown of the order parameter \cite{RevModPhys.49.435}. Effectively, the fluctuations of the order parameter contribute an additional almost gapless mode, with relaxation time $\tau_{o.p.}$ that can become arbitrarily large close to criticality. Therefore, in our system, we have two independent small scales $\frac{1}{\tau_{hydro}}$ and $\frac{1}{\tau_{o.p.}}$ corresponding to different physics. The relaxation time $\tau_{hydro}$ is controlled by the scale $l_{inh.}$, whereas $\tau_{o.p.}$ is controlled by the proximity to the critical point. We will therefore need two independent small dimensionless parameters $\lambda$ and $\ves$ to organise our expansion.

For the case of superfluids and in the absence of explicit symmetry breaking, we assume that there is a critical point in the thermodynamic plane $(T,\mu)$, with critical temperature $T_c(\mu)$. For the purposes of our paper, we will be interested in a family of nearly critical configurations at temperature and chemical potential along a curve $(T(\ves),\mu(\ves))$. The small dimensionless parameter $\ves$ is chosen such that
\begin{align}\label{eq:mu_T}
    T(\ves)=T_c(\mu)+\mathcal{O}(\ves^2)\,,\quad  \mu(\ves)=\mu+\mathcal{O}(\ves^2)\,.
\end{align}
As we have argued, in order to describe the full dynamics of our system in this region, we need to incorporate the modulus  $|\psi|$. This is in addition to the standard hydrodynamic variables $\mu,T,u^\mu$ and the phase of $\psi$, which are sufficient for the description of superfluids away from criticality.

On the other hand, the small parameter $\lambda$ that we introduced earlier keeps track of the number of derivatives acting on our local fields. In this way, we consider the following scaling properties for the out-of-equilibrium effective variables,
\begin{align}
&\mu,\,T,\,u^\mu\sim1,\quad \psi\sim\ves\,,\quad s_\psi\sim\lambda\, \ves\,,\quad \mathcal{F}_\psi \sim\ves^3+\lambda\, \ves\,,\nn
    &\partial_\mu\mu,\,\partial_\mu T,\,\nabla_\mu u^\nu,\, F_{\mu\nu}\sim\lambda\, \ves^2\,,\quad \hat{D}_\mu\psi\sim\lambda\, \ves,\quad \hat{D}_\mu s_\psi\sim\lambda^2\,\ves\,.
\end{align}
The $\ves$ part is dictated by mean field theory arguments. In addition, based on \eqref{eq:mu_T}, we have assumed that the fluctuating part of $\mu,\,T,\,u^\mu$ is of order $\ves^2$.

So far in the discussion, the scales $\lambda$ and $\ves$ have been formally and conceptually independent. However, in order to actually \emph{solve} the equations of motion of the theory, one has to choose the relative order of magnitude between these two. It is straightforward to see that in the double $\lambda,\,\ves$ expansion, the leading terms we can write in the equation of motion for the order parameter are $\hat{D}_u\psi\sim\lambda\,\ves $  and   $\mathcal{F}_\psi\sim\ves^3$. Thus, if we choose $\lambda\ll\ves^2$, then to leading order $\mathcal{F}_\psi=0$, and the modulus of the order parameter decouples. In this limit, $|\psi|$ is effectively no longer a slow degree of freedom and we can integrate it out by solving its equation of motion order by order in gradients. As we will later see, in this limit we recover superfluid hydrodynamics. In the opposite limit, $\lambda\gg\ves^2$, we are effectively placing the system closer and closer to the critical point, and we should see traits of the normal charged fluid emerging. We will examine the two limits in Section \ref{sec:Asymptotic}. In the more interesting scaling region $\lambda\sim\ves^2$, the dynamics of the order parameter and the normal fluid remain coupled and, for this reason, we will assume this relative scaling for the rest of this section.

It is well known that away from equilibrium, the local hydrodynamic variables $T,\,\mu,\,u^\mu$ are ambiguous\cite{Kovtun:2012rj} since we are allowed to perform local field redefinitions of the form,
\begin{align}
    T \to T+\delta T,\, \mu\to\mu+\delta\mu,\,u^\mu\to u^\mu+\delta u^\mu\,,
\end{align}
with $\delta T,\delta\mu,\,\delta u^\mu$ being of order $\mathcal{O}(\lambda\,\ves^2 )$.
A standard choice that we found convenient for our purposes is to fix this freedom by imposing the transverse frame conditions,
\begin{align}\label{eq:trans_frame_cond}
    T^{\mu\nu}_{diss}u_\nu=0,\quad J^\mu_{diss}u_\mu=0\,.
\end{align}

We will now turn our attention to the expansion of the dissipative terms of equation \eqref{eq:diss_mod}. In the expansion scheme that we have introduced above, and in the scaling region $\lambda\sim\ves^2$, we would like to retain only the leading corrections in $T^{\mu\nu}_{diss},\,J^\mu_{diss}$, of order $\mathcal{O}\left(\lambda\,\ves^2,\ves^4\right)$, while in the expression for $E_{diss}$ we want to  include the leading ($\mathcal{O}(\ves^3,\lambda\,\ves)$) and next-to-leading contributions ($\mathcal{O}(\ves^5,\lambda^2 \ves,\lambda\,\ves^3)$).

As always in hydrodynamics, when writing dissipative corrections, we can take into account the leading equations of motion and eliminate certain derivative correction terms in terms of others. To leading order, the continuity equations \eqref{eq:contin} yield,
\begin{align}\label{eq:lead_cont}
    \partial_\mu\varrho\, \partial_u\mu+\partial_T\varrho\, \partial_uT+\partial_{|\psi|^2 }\varrho\,\partial_u|\psi|^2+\varrho\, \nabla_\mu u^\mu+2 q_e \mathrm{Im}\left(\psi^\ast\mathcal{F}_\psi\right)=\mathcal{O}(\lambda^2 \ves^2)\,,\nn
    \partial_\mu\epsilon\, \partial_u\mu+\partial_T\epsilon\, \partial_uT+\partial_{|\psi|^2 }\epsilon\,\partial_u|\psi|^2+(\epsilon+p)\, \nabla_\mu u^\mu+2q_e\mu\,\mathrm{Im}\left(\psi^\ast\mathcal{F}_\psi\right)=\mathcal{O}(\lambda^2 \ves^2)\,.
\end{align}
After introducing the complex coefficient $\Gamma_0$, the equation of motion for the order parameter gives,
\begin{align}\label{eq:lead_op}
\hat{D}_u\psi=-2\overline{\Gamma}_0\,\mathcal{F}_\psi+\mathcal{O}(\ves^5)\,.
\end{align}
Using the leading order equations of motion, we can trade $\partial_u\mu$, $\partial_u T$, $\partial_u|\psi|^2$ for $\nabla_\mu u^\mu$ and $\psi^\ast \mathcal{F}_\psi$.

To the order that we have specified above, the most general expressions we can write in the transverse frame are\footnote{In writing the above we have already used the known constraints for normal fluids to relate e.g. the coefficients of $\nabla^{\perp}_\mu\mu,\,\nabla^{\perp}_\mu T,\,P^{\mu\nu}F_{\nu\rho}u^\rho$ in $J^\mu_{diss}$. We have also defined the shear tensor in the usual way, $ \sigma^{\mu\nu}=P^{\mu\rho}P^{\nu\sigma}\left(2\,\nabla_{(\rho}u_{\sigma)}-\frac{2}{d-1} g_{\rho\sigma}\nabla_\tau u^\tau\right)$. },
\begin{align}\label{eq:corr_constit}
    T^{\mu\nu}_{diss}=&-\eta\,\sigma^{\mu\nu}-Z_1\, P^{\mu\nu}\nabla_\rho u^\rho-2\,\mathrm{Re}\left[Z_3\, \mathcal{F}_\psi\,\psi^\ast\right]P^{\mu\nu}\,,\nn
    J^\mu_{diss}=&-T \sigma  P^{\mu\nu} \left( \nabla_\nu\left(\frac{\mu}{T}\right)-\frac{F_{\nu\rho}\,u^\rho}{T}\right)\,,\nn
    E_{diss}=&-2\,\overline{\Gamma}_0\,\mathcal{F}_\psi-Z_n\, \psi^2\, \mathcal{F}_\psi^\ast+Z_2\,\psi\, \nabla_\mu u^\mu-Z_\pi\,  \hat{D}_u \mathcal{F}_\psi\,,
\end{align}
where we made the assumption that all the transport coefficients remain finite close to the critical point. In particular, the coefficients $\eta,\, Z_1,\,\sigma$ are the equivalents of the shear viscosity, bulk viscosity, and conductivity of normal fluids. The new coefficients $Z_3,\,\Gamma_0, Z_2,\,Z_n$ and $Z_\pi$ are intimately related to the existence of the condensate, and can be complex, in general. Moreover, we can assume that all of our coefficients are analytic functions of $\mu, T, |\psi|^2$ and therefore expandable in powers of $\ves^2$.

At the same order in $\ves$, we could have also written a term $\mathrm{Re}\left(\psi^\ast D^{\mu \perp}\psi \right)$ in $J^\mu_{diss}$, and a term $(D_\mu^{\perp}
)^2\psi$ in $E_{diss}$. However, in the end, we would have to set both to zero, either by invoking the positivity of entropy production or by noticing that none of these have to be zero in equilibrium and are thus non-dissipative\footnote{Alternatively, we can impose Onsager reciprocity for $\lambda\sim\ves^2$ and in the hydrodynamic regime, for $\lambda\ll\ves^2$. This also sets the coefficients of these two terms to zero.}. The other two possible terms we could have naively added are a second time-derivative term $( \hat{D}_u)^2\psi$, or a $\hat{D}_u s_\psi$ term in $E_{diss}$. However, both of them can be traded for a term $ \hat{D}_u \mathcal{F}_\psi$ at this order, using \eqref{eq:lead_cont} and \eqref{eq:lead_op}.

After examining the form of the individual terms that can enter our constitutive relations and equations of motion, we will examine the constraints that our transport coefficients in \eqref{eq:corr_constit} have to satisfy. The first general requirement that we would like to examine is Onsager reciprocity \cite{Onsager:1931jfa,Onsager:1931kxm}, which leads to the relations\footnote{The $Z_3$ term of $T^{\mu\nu}_{diss}$ and the $Z_n$ term of $E_{diss}$ in \eqref{eq:corr_constit}, given \eqref{eq:F_wrt_psi}, include contributions with two spatial derivatives of the condensate. These terms, for $\lambda\sim\ves^2$, are of order $\ves^6$ and $\ves^7$ respectively, i.e. outside the regime of validity of \eqref{eq:corr_constit}. However, in the superfluid regime $\lambda\ll\ves^2$, these terms  lead to first order corrections and must be kept, so that Onsager reciprocity is satisfied in this regime as well. See also Subsection \ref{sec:Integr_out} for more details.}
\begin{align}\label{eq:Recipr}
    Z_3=Z_2\,,\quad\mathrm{Im}(Z_n)=0\,.
\end{align}

Next, we turn our attention to the entropy current for our theory. In principle, in the presence of dissipative effects, one should consider a generic entropy current, whose expression may deviate from the equilibrium form \eqref{eq:entr_can}, due to derivative corrections (see, for example, \cite{Bhattacharya:2011tra}, \cite{Son:2009tf}, \cite{Bhattacharyya:2012nq}). We can express the entropy current as, 
\begin{align}\label{eq:entr_current}
    s^\mu=\frac{p}{T}u^\mu-T^{\mu\nu}\frac{u_\nu}{T}-\frac{\mu}{T}J^\mu+\mathrm{Re}\left(\frac{w_0}{T} \hat{D}_u\psi\,D^{\perp\mu}\psi^\ast\right)+\Delta s^\mu\,,
\end{align}
 where $\Delta s^\mu$ includes other possible derivative corrections. In the scaling region $\lambda\sim\ves^2$ we will need to find the entropy production $\nabla_\mu s^\mu$ up to order $\mathcal{O}(\ves^8)$, hence we will need $s^\mu$ up to order $\mathcal{O}(\ves^6)\,.$ Contrary to hydrodynamics, the Keldysh-Schwinger formalism provides us with a specific form for the entropy current, at each order in the $\lambda$ expansion. Our analysis in Section \ref{sec:KS_hydro} will dictate that $\Delta s^\mu=\mathcal{O}(\lambda^2\ves^2)$, i.e. it will include terms with at least two derivatives. It is not difficult to check that possible contributions to $\Delta s^\mu$ at this order are then either terms with second derivatives of the hydrodynamic variables (or of the sources $g_{\mu\nu},\,A_\mu$), terms with a product of first derivatives of the condensate or, finally, terms with second derivatives of the condensate.   

We must now impose that the local entropy of our system can only increase in time. As usual, this requirement will constrain the various terms that would otherwise be allowed in our hydrodynamic expansion. We can check that, on-shell,
\begin{align}\label{eq:entropy_prod}
   & \nabla_\mu s^\mu=
    -\frac{2}{T}\,\mathrm{Re}\left(\mathcal{F}_\psi^\ast\,E_{diss}\right)-\frac{\nabla_\mu u_\nu}{T}T^{\mu\nu}_{diss}-\left(\nabla_\mu\left(\frac{\mu}{T}\right)-\frac{F_{\mu\nu}u^\nu}{T}\right)J^\mu_{diss}+\nabla_\mu\Delta s^\mu=\nn 
    &\frac{2}{T|\psi|^2}\left(2\,\mathrm{Re}(\Gamma_0)+\mathrm{Re}(Z_n)|\psi|^2\right)\left(\mathrm{Re}(\mathcal{F}_\psi^\ast\, \psi)\right)^2+\frac{2}{T|\psi|^2}\left(2\,\mathrm{Re}(\Gamma_0)-\mathrm{Re}(Z_n)|\psi|^2\right)\left(\mathrm{Im}(\mathcal{F}_\psi^\ast\, \psi)\right)^2\nn
    &+\frac{4}{T}\mathrm{Im}(Z_2)\nabla_\mu u^\mu\, \mathrm{Im}(\mathcal{F}_\psi^\ast\, \psi)+\frac{Z_1}{T}\left(\nabla_\mu u^\mu\right)^2+\frac{\eta}{T}\left(\sigma_{\mu\nu}\right)^2+T\sigma \left(\nabla_\mu^\perp\left(\frac{\mu}{T}\right)-\frac{F_{\mu\nu}u^\nu}{T}\right)^2\nn&-\frac{2}{T}\mathrm{Re}(Z_\pi\,\mathcal{F}_\psi^\ast\,\hat{D}_u\mathcal{F}_\psi)+\nabla_\mu \Delta s^\mu\,,
\end{align}
with $\mathcal{F}_\psi^\ast$ as given in \eqref{eq:F_wrt_psi}.
As we already mentioned above, in the scaling regime $\lambda\sim\ves^2$ \eqref{eq:entropy_prod} includes terms up to order $\mathcal{O}(\ves^8)$. We note that the contributions in the last line of \eqref{eq:entropy_prod} disentangle from the rest of the contributions, which constitute a quadratic form on their own\footnote{The quadratic form is evaluated on a vector with components $\nabla_\mu u^\mu$, $\mathrm{Re}(\mathcal{F}_\psi^\ast \psi)$, $\mathrm{Im}(\mathcal{F}_\psi^\ast \psi)$, $\sigma_{\mu\nu}$ and $\nabla_\mu^\perp(\frac{\mu}{T})-\frac{F_{\mu\nu}u^\nu}{T}$.}. Demanding that this quadratic form is semi-positive definite, we find,
\begin{align}\label{eq:Ineqs_1}
    \eta&\geq0\,,\sigma\geq0,\,Z_1\geq 0\,,
\end{align}
as in normal fluid hydrodynamics. Moreover, we discover the inequalities,
\begin{align}
&2\,\mathrm{Re}\left(\Gamma_0\right)+\mathrm{Re}
(Z_n)|\psi|^2\geq 0\,,\nn \label{eq:Ineqs_2}
    & 2\,\mathrm{Re}\left(\Gamma_0\right)-\mathrm{Re}
(Z_n)|\psi|^2\geq 0\,, \nn
   &2|\psi|^2\,\mathrm{Im}(Z_2)^2  \leq  \left(2\,\mathrm{Re}(\Gamma_0)-\mathrm{Re}(Z_n)|\psi|^2\right)Z_1 \,.
\end{align}
The last two lines of \eqref{eq:entropy_prod} can only lead to inequalities which will involve $Z_\pi$ and the coefficients appearing in $\Delta s^\mu$, which we shall not investigate further. In order to check the semipositivity of these terms, we would need to consider even higher order terms in \eqref{eq:corr_constit}. 

We should point out that the coefficient $\mathrm{Re}(Z_2)$ cancels out in the entropy production expression, playing a role similar to the ``compressibility'' $A_\phi$ of \cite{Stephanov:2017ghc}. In particular, $\mathrm{Re}(Z_2)$ multiplies the projection operator $P^{\mu\nu}$ in the stress tensor constitutive relation \eqref{eq:corr_constit} and can thus be viewed as a non-equilibrium correction to pressure. As an aside, the authors of \cite{Stephanov:2017ghc} introduce a generalised (partial-equilibrium) pressure $p_{(+)}$, which in our notation can be expressed as $p_{(+)}=p-2\,\mathrm{Re}(Z_2)\,\mathrm{Re}\left(\mathcal{F}_\psi\,\psi^\ast\right)$. However, we are not going to make use of this quantity in the rest of this paper.

The dissipative corrections \eqref{eq:corr_constit}, along with the constraints \eqref{eq:Recipr}, \eqref{eq:Ineqs_1}, \eqref{eq:Ineqs_2}, are the main results of this section. With the exception of the extra term coming with the  coefficient $Z_\pi$ in the order parameter equation, these results are a covariant generalisation of the work of Khalatnikov and Lebedev \cite{KhalatnikovLebedev1978}, with which we find agreement in the appropriate non-relativistic limit. The correspondence between the transport coefficients of \eqref{eq:corr_constit} and \cite{KhalatnikovLebedev1978} is\footnote{The coefficients $\zeta_1,\,\zeta_2,\,\zeta_3$ of \cite{KhalatnikovLebedev1978} are not related to the superfluid bulk viscosities we introduce in Subsection \ref{sec:Integr_out}.}
\begin{align}
    2\overline{\Gamma}_0\leftrightarrow\zeta_1,\quad Z_n\leftrightarrow\zeta_2,\quad Z_2\leftrightarrow-\zeta_3,\quad Z_1\leftrightarrow\zeta_5,\quad\eta\leftrightarrow\zeta_7,\quad\sigma\leftrightarrow\zeta_6,\quad Z_\pi \leftrightarrow0\,.
\end{align}
In Appendix \ref{app:Elim_Z_pi} , we discuss in more detail the necessity of the $Z_\pi$ term in our theory.

\subsection{Effective theory at the linear level}\label{sec:Eff_linear}

In this subsection, we will consider the linearisation of the effective theory around an equilibrium thermodynamic state with the external sources turned off. More concretely, we will parametrise our hydrodynamic variables according to,
\begin{align}
   \mu&=\mu_0+\delta\mu,\quad T=T_0+\delta T,\quad u^\mu_0=\delta^\mu_t,\quad\delta u^t=\frac{\delta g_{tt}}{2},\quad \delta u^i=\delta v^i-\delta g_{it}\,,\nn
   \psi&=\rho_v+\delta\rho_v+i q_e\, \rho_v\, \delta\theta\,,
\end{align}
with $T_0,\mu_0,\rho_v$ constants. Similarly, for the external sources we will write,
\begin{align}
g_{\mu\nu}=\eta_{\mu\nu}+\delta g_{\mu\nu}\,,\quad
A=(\mu_0+\delta A_t)\,dt+\delta A_i\, dx^i\,,\quad s_\psi=\delta s_\psi^R+i\, \delta s_\psi^I\,,
\end{align}
with $\eta_{\mu\nu}$ the Minkowski metric and it also useful to define the gauge invariant combination $\delta m_\mu=\partial_\mu \delta \theta+\delta A_\mu$\footnote{We use the conventions: $\delta g^{\mu\nu}\equiv-\eta^{\mu\rho}\eta^{\nu\sigma}\delta g_{\rho\sigma}$, $\delta m^i\equiv\delta^{ij}\delta m_j$.}. As usual, we will assume that all linear fluctuations have a spacetime dependence $\sim e^{-i\omega t+i k_i x^i}$, with $\omega$ the frequency and $k_i$ the wavevector of the perturbation.

Similarly to \cite{Stephanov:2017ghc}, we have found it useful to define the variable $\pi$,
\begin{align}\label{eq:pi_def}
    \pi=\left(\frac{\partial f}{\partial |\psi|}\right)_{\mu,T}\,,
\end{align}
with $f$ as defined in equation \eqref{eq:f_tot_def}. This variable is conjugate to $|\psi|$, and in thermodynamic equilibrium it vanishes in the absence of external sources. Notice that this condition for the background determines $\rho_v$ as a function of $\mu_0, T_0$.

As we discuss in Appendix \ref{app:susc_def}, after a Legendre transformation of the thermodynamic potential $f$, all thermodynamic quantities can be viewed as functions of $\mu$, $T$ and $\pi$. In this ensemble, we can introduce susceptibilities through,
\begin{align}
\delta\varrho&=\chi \,\delta\mu+\xi\,\delta T+\nu_{\mu\rho}\,\delta\pi\,,\nn
\delta s&=\xi\,\delta\mu+\frac{c_\mu}{T_0}\,\delta T+\nu_{T\rho}\,\delta\pi\,,\nn 
\delta |\psi|&=\nu_{\mu\rho}\,\delta\mu+\nu_{T\rho}\,\delta T+\nu_{\rho\rho}\,\delta\pi\,.
\end{align}
with all of them being evaluated at $\mu=\mu_0,\,T=T_0,\,\rho_v=\rho_v(\mu_0,T_0)$. 
Similarly, for the fluctuations of pressure and energy, we can write,
\begin{align}
    \delta p=\varrho_0\, \delta\mu+s_0\,\delta T+\rho_v\, \delta s_\psi^R\,,\quad
    \delta \epsilon=T_0\, \delta s+\mu_0\, \delta\varrho-\rho_v\,\delta s_\psi^R\,,
\end{align}
since $\pi_0$=0 in the background.  

The definitions above allow us to express the fluctuations of the stress tensor and conserved current according to,
\begin{align}\label{eq:const_Linear}
    \delta T^{tt}&=\left(c_\mu +\mu_0\,\xi\right)\delta T+\left(\mu_0\, \chi+T_0\, \xi\right)\delta\mu- \epsilon_0 \, \delta g^{tt}+\left(\mu_0\, \nu_{\mu\rho}+T_0\, \nu_{T \rho}\right)\delta\pi-\rho_v\,\delta s_\psi^R\,,\nn
    \delta T^{ti}&=(\epsilon_0+p_0-\mu_0^2\, \chi_{JJ})\delta v^i-\epsilon_0\, \delta g^{ti}-\mu_0\,\chi_{JJ}\,\delta m^i\,,\nn
    \delta T^{ij}&=p_0\, \delta g^{ij}+\left(s_0\,\delta T+\varrho_0\,\delta\mu+\rho_v\, \delta s_\psi^R\right)\delta^{ij}- Z_1\, \delta^{ij}\left(\partial_k(\delta v^k-\delta g^{kt})+\partial_t \delta g_{kl}\,\frac{\delta^{kl}}{2}\right)\nn&
    -\eta\,\delta \sigma^{ij}-2\,\rho_v\,\mathrm{Re}\left[Z_2\, \delta \mathcal{F}_\psi\right]\delta^{ij}\,,\nn
    \delta J^t&=\xi\,\delta T+\chi\,\delta\mu+\nu_{\mu\rho}\,\delta\pi-\varrho_0\, \frac{\delta g^{tt}}{2}\,,\nn 
    \delta J^i&=\left(\varrho_0-\mu_0\, \chi_{JJ}\right)\,\delta v^i-\chi_{JJ}\,\delta m^i-\varrho_0\,\delta g^{ti}-\sigma\,\delta^{ij}\left(\partial_j \delta\mu- \frac{\mu_0}{T_0}\,\partial_j\delta T-\delta F_{jt}\right)\,,
\end{align}
with,
\begin{align}
    \delta \mathcal{F}_\psi&=-\frac{w_0}{2}\left(\partial_i^2\delta\rho_v+i q_e \rho_v \partial_i(\delta m^i+\mu_0\, \delta v^i)\right)+\frac{\delta\pi}{2}-\frac{\delta s_\psi^R+i \delta s_\psi^I}{2}\,,\nn
    \delta\sigma^{ij}&=2\,\partial^{(i}\delta v^{j)}-2\,\partial^{(i}\delta g^{j)t}-\partial_t\delta g^{ij}-\delta^{ij}\left(\partial_k(\delta v^k-\delta g^{kt})+\partial_t \delta g_{kl}\,\frac{\delta^{kl}}{2}\right)\,.
\end{align}
where we have also made use of the identification \eqref{eq:cur_cur_susc} for the current-current susceptibility $\chi_{JJ}$. Finally, after expanding the equation of motion for the order parameter in \eqref{eq:diss_mod} and the dissipative corrections in \eqref{eq:corr_constit}, we find,
\begin{align}
&\partial_t \delta \rho_v+i q_e \rho_v \delta m_t+iq_e\,\mu_0\,\rho_v\,\frac{\delta g_{tt}}{2}-iq_e\rho_v\, \delta \mu=\nn& \label{eq:ord_linear}-2\, \overline{\Gamma}_0\, \delta \mathcal{F}_\psi-Z_n\,\rho_v^2\, \delta \mathcal{F}_\psi^\ast
+Z_2\,\rho_v\left(\partial_k(\delta v^k-\delta g^{kt})+\partial_t \delta g_{kl}\,\frac{\delta^{kl}}{2}\right)-Z_\pi\, \partial_t\,\delta \mathcal{F}_\psi.
\end{align}

Let us momentarily focus on a conformally invariant theory, for which the stress tensor has to be traceless. In that case, the tracelessness of the linearised stress tensor, given above in \eqref{eq:const_Linear} leads to,
\begin{align}\label{eq:conform_inv_1}
    Z_1=0,\quad \mathrm{Im}\left(Z_2\right)=0,\quad\mathrm{Re}(Z_2)=-\frac{\mu_0\, \nu_{\mu\rho}+T_0\,\nu_{T\rho}}{\rho_v(d-1)}\,.
\end{align}
Assuming that the scaling dimension of $\rho_v$ is $\Delta_\psi$, then $\rho_v=T^{\Delta_\psi}f_\rho(\frac{\mu}{T})$ for some arbitrary function $f_\rho$. Using this relation, the last condition in \eqref{eq:conform_inv_1} simplifies to
\begin{align}\label{eq:conform_inv_2}
    \mathrm{Re}(Z_2)=-\frac{\Delta_\psi}{d-1}\,.
\end{align}

For later convenience, it will also be useful to define susceptibilities in the fixed $s$, $\varrho$, $\pi$ ensemble and relate them to susceptibilities of the fixed $T$, $\mu$ and $\pi$ ensemble. To do this, we can use the chain rule to show that the partial derivatives of any thermodynamic quantity $A$ in the two ensembles are related according to,
\begin{align}\label{eq:ch_ens}
    \partial_T (A)_{\mu,\pi}&=\frac{c_\mu}{T}\,\partial_s (A)_{\varrho,\pi}+\xi\, \partial_\varrho(A)_{s,\pi}\,,\nn
    \partial_\mu (A)_{T,\pi}&=\xi\,\partial_s (A)_{\varrho,\pi}+\chi\, \partial_\varrho(A)_{s,\pi}\,,\nn
    \partial_{\pi} (A)_{\mu,T}&=\nu_{T\rho}\,\partial_s (A)_{\varrho,\pi}+\nu_{\mu\rho}\, \partial_\varrho(A)_{s,\pi}+\partial_{\pi}(A)_{\varrho,s}\,.
\end{align}
In particular, we have found useful the introduction of the following susceptibilities involving the amplitude,
\begin{align}
    \nu_{\rho s}=\partial_s (\rho_v)_{\varrho,\pi}\,,\quad\nu_{\rho \varrho}=\partial_\varrho(\rho_v)_{s,\pi}\,,\quad \tilde{\nu}_{\rho \rho}=\partial_{\pi} (\rho_v)_{\varrho,s}\,.
\end{align}

We conclude this section by noting that the scaling of susceptibilities and background thermodynamic quantities with the small parameter $\ves$ can be predicted in mean field theory, as we discuss in Appendix \ref{app:mean_exp}.

\section{Asymptotic regions of the effective theory}\label{sec:Asymptotic}

In this section, we will expand on the discussion regarding the relative scaling between the expansion parameters $\lambda$ and $\ves$ of Subsection \ref{sec:Dissipation}. In the context of the linearised theory presented in the previous section, we will investigate the two asymptotic regions of wavevector $k$ magnitudes in our effective theory. More specifically, Subsection \ref{sec:Integr_out} considers $|k|$ to be the smallest scale in our problem, and, as we argued before, we recover the conventional superfluid hydrodynamics with fixed transport coefficients. In Subsection \ref{sec:Normal_fluid} we confirm that for large $|k|$, we recover the hydrodynamics of the normal phase.

\subsection{Superfluid hydrodynamics}\label{sec:Integr_out}

First, we examine the limit of small momenta and frequencies for linearised perturbations in our system, obeying equations \eqref{eq:const_Linear}, \eqref{eq:ord_linear}. Hence, we would like to focus on energy scales much smaller than the gap of the amplitude mode. This is, by definition, the regime of validity of conventional superfluid hydrodynamics \cite{Donos:2022www, Herzog:2011ec,Bhattacharya:2011tra}, far away from the critical point. 

In order to take the low-energy limit systematically, we reintroduce the expansion parameter $\lambda$, as in Subsection \ref{sec:Dissipation}, which sets the scale for the wavevector, the frequency, and the external gauge field and metric, so that\footnote{Note that we will set $\delta s_\psi^R=\delta s_\psi^I=0$ for this computation. } 
\begin{align}\label{eq:lambda_scaling_1}
k_i,\,\omega,\,\delta g_{\mu\nu}\,,\delta A_\mu=\mathcal{O}(\lambda)\,.
\end{align}
The hydrodynamic variables are also expandable in $\lambda$, according to,
\begin{align}
    \delta T&=\delta T_1\, \lambda+\delta T_2\, \lambda^2+\cdots\,,\quad 
     \delta \mu=\delta \mu_1\, \lambda+\delta \mu_2\, \lambda^2+\cdots\,,\quad
      \delta v^i=\delta v^i_1\, \lambda+\delta v^i_2\, \lambda^2+\cdots\,,\nn
       \delta \theta&=\delta \theta_0+\delta \theta_1\, \lambda+\cdots\,,\quad
        \delta \rho_{v}=\delta \rho_{v,1} \lambda+\delta \rho_{v,2}\, \lambda^2+\cdots\,.
\end{align}
Solving the order parameter equation of motion \eqref{eq:ord_linear} up to order $\lambda^2$, we can integrate out the amplitude to find,
\begin{align}\label{eq:sol_ampl}
    \delta \rho_{v,1}=&\nu_{T\rho}\,\delta T_1+\nu_{\mu\rho}\,\delta\mu_1\,,\nn
    \delta \rho_{v,2}=&\nu_{T\rho}\,\delta T_2+\nu_{\mu\rho}\,\delta\mu_2+\frac{\nu_{\rho\rho}}{\mathrm{Re}\Gamma_0+\frac{\rho_v^2}{2}\,\mathrm{Re}Z_n}\biggl(\frac{ \mathrm{Im}\Gamma_0\,\chi_{JJ}}{ q_e\,\rho_v}i k_i\left(\delta m^i_1+\mu_0\, \delta v^i_1\right)+\nn &   i \omega\left(\nu_{T\rho}\,\delta T_1+\nu_{\mu\rho}\,\delta\mu_1\right)
     +\rho_v\, \mathrm{Re}Z_2\left(ik_i\left(\delta v^i_1-\delta g^{it}\right)-i\omega\, \delta g_{ij}\frac{\delta^{ij}}{2}\right)\biggl)\,.
\end{align}

The above results can be thought of as the constitutive relations for the amplitude in a conventional hydrodynamic expansion. The leading order result, $\delta\rho_{v,1}$, is exactly what one would expect from thermodynamics, whereas $\delta\rho_{v,2}$ gives the leading correction, at first order in derivatives of the hydrodynamic variables. It is important to note that the result is local in derivatives. This is in accordance with the fact that the amplitude in the small $k$ limit (equivalently, large gap limit) is a UV degree of freedom, and so integrating it out should not break locality of the theory.

The next step is to expand the stress tensor and current in equation \eqref{eq:const_Linear}, as well as the imaginary part of \eqref{eq:ord_linear} up to order $\lambda^2$ and eliminate the amplitude using \eqref{eq:sol_ampl}. After eliminating $\delta\rho_{v,2}$ in the stress tensor and electric current, new terms with time derivatives of $\delta T,\delta\mu$ appear. These can be traded for spatial derivatives of the fluid velocity after using the ideal level continuity equations. At the same time, we notice that we need to perform a simple redefinition of the local temperature and chemical potential in order to preserve the transverse frame choice \eqref{eq:trans_frame_cond}.

After the above manipulations, the linearised constitutive relations in coordinate space take the standard form in the transverse frame \cite{Herzog:2011ec},
\begin{align}\label{eq:conv_super}
    \delta T^{tt}&=\left(c_\mu+\mu_0\,\xi\right)\delta T+\left(\mu_0\,\chi+T_0\,\xi\right)\delta\mu-\epsilon_0\,\delta g^{tt}\,,\nn
    \delta T^{ti}&=(\epsilon_0+p_0-\mu_0^2\,\chi_{JJ})\delta v^i-\epsilon_0\, \delta g^{ti}-\mu_0\,\chi_{JJ}\,\delta m^i\,,\nn
     \delta T^{ij}&=p_0\, \delta g^{ij}+\left(s_0\,\delta T+\varrho_0\,\delta\mu\right)\delta^{ij}- \zeta_1\, \delta^{ij}\left(\partial_k(\delta v^k-\delta g^{kt})+\partial_t \delta g_{kl}\,\frac{\delta_{kl}}{2}\right)\nn&
    -\eta\,\delta\sigma^{ij}
+\chi_{JJ}\,\zeta_2\,\partial_i\left(\delta m^i+\mu_0\, \delta v^i\right)\,,\nn
    \delta J^t&=\xi\,\delta T+\chi\, \delta\mu-\varrho_0\,\frac{\delta g^{tt}}{2}\,,\nn
    \delta J^i&=\left(\varrho_0-\mu_0\,\chi_{JJ}\right)\delta v^i-\chi_{JJ}\,\delta m^i-\varrho_0\,\delta g^{ti}-\sigma \,\delta^{ij}\left(\partial_j \delta\mu- \frac{\mu_0}{T_0}\,\partial_j\delta T-\delta F_{jt}\right)\,.
\end{align}
From the expansion of the imaginary part of the equation of motion of the complex scalar \eqref{eq:ord_linear} we obtain the Josephson relation,
\begin{align}\label{eq:Jos_rel}
    \delta\mu=\delta m_t+\mu_0\frac{\delta g_{tt}}{2}+\zeta_2\left(\partial_k(\delta v^k-\delta g^{kt})+\partial_t \delta g_{kl}\,\frac{\delta_{kl}}{2}\right)-\chi_{JJ}\,\zeta_3\,\partial_i\left(\delta m^i+\mu_0\, \delta v^i\right)\,.
\end{align}
The three bulk viscosities $\zeta_i$ introduced above as expressed in terms of the transport coefficients of the nearly critical theory are given by,
\begin{align}\label{eq:bulk_visc_conv}
\zeta_1&=\frac{\left(s\,\nu_{\rho s}+\varrho\, \nu_{\rho \varrho}+\rho_v\, \mathrm{Re}Z_2\right)^2}{\mathrm{Re}\Gamma_0+\frac{\rho_v^2}{2}\,\mathrm{Re}Z_n}+Z_1\,,\nn
   \zeta_2&=\frac{\left(s\,\nu_{\rho s}+\varrho\, \nu_{\rho \varrho}+\rho_v\, \mathrm{Re}Z_2\right)}{q_e\,\rho_v \left(\mathrm{Re}\Gamma_0+\frac{\rho_v^2}{2}\,\mathrm{Re}Z_n\right)}\left(-\mathrm{Im}\Gamma_0+q_e\,\rho_v\, \nu_{\rho\varrho}\right)-\frac{\mathrm{Im}Z_2}{q_e}\,,\nn
   \zeta_3&=\frac{\left(-\mathrm{Im}\Gamma_0+q_e\,\rho_v\, \nu_{\rho\varrho}\right)^2}{q_e^2\,\rho_v^2\,\left(\mathrm{Re}\Gamma_0+\frac{\rho_v^2}{2}\,\mathrm{Re}Z_n\right)}+\frac{\mathrm{Re}\Gamma_0-\frac{\rho_v^2}{2}\,\mathrm{Re}Z_n}{q_e^2\,\rho_v^2}\,,
\end{align}
in the fixed $s,\varrho,\pi$ ensemble.

After using the ideal equations of motion and fixing the hydrodynamic frame, the constitutive relation for the amplitude \eqref{eq:sol_ampl} becomes,
\begin{align}
\delta\rho_v=\nu_{T\rho}\,\delta T+\nu_{\mu\rho}\,\delta\mu+\zeta_{\rho,v}\left(\partial_k(\delta v^k-\delta g^{kt})+\partial_t \delta g_{kl}\,\frac{\delta_{kl}}{2}\right)+\zeta_{\rho,c}\,\partial_i\left(\delta m^i+\mu_0\, \delta v^i\right)\,,
\end{align}
with,
\begin{align}
    \zeta_{\rho,v}=\tilde{\nu}_{\rho\rho}\frac{\left(s\,\nu_{\rho s}+\varrho\, \nu_{\rho \varrho}+\rho_v\, \mathrm{Re}Z_2\right)}{\mathrm{Re}\Gamma_0+\frac{\rho_v^2}{2}\,\mathrm{Re}Z_n}\,,\quad
    \zeta_{\rho,c}=\frac{\tilde{\nu}_{\rho\rho}\,\chi_{JJ}}{\mathrm{Re}\Gamma_0+\frac{\rho_v^2}{2}\,\mathrm{Re}Z_n}\left(\frac{\mathrm{Im}\Gamma_0}{q_e\, \rho_v}-\nu_{\rho\varrho}\right)\,.
\end{align}
We note that this relation defines a new set of transport coefficients, $\zeta_{\rho,v},\,\zeta_{\rho,c}$ for the superfluid, which are not present in the usual stress tensor and current constitutive relations.

Equations \eqref{eq:conv_super} are \eqref{eq:Jos_rel} are indeed the constitutive relations of first-order superfluid hydrodynamics in the transverse frame, when linearised around a static homogeneous and isotropic background. The requirement of positive entropy production for standard superfluids \cite{Herzog:2011ec,Donos:2022www} constrains the bulk viscosities to satisfy  $\zeta_1\geq 0$, $\zeta_3\geq 0$ and $|\zeta_2|\leq \sqrt{\zeta_1}\sqrt{\zeta_3}$. It is important to point out that these immediately follow from our inequalities \eqref{eq:Ineqs_1},\eqref{eq:Ineqs_2}. This provides further evidence about the correctness of the entropy current expression of equation \eqref{eq:entr_current}.

For a conformal superfluid, the tracelessness of the stress tensor dictates that $\zeta_1=\zeta_2=0$\cite{Herzog:2011ec,Donos:2022www}. The expressions \eqref{eq:bulk_visc_conv} for $\zeta_1,\,\zeta_2$ and the conditions \eqref{eq:conform_inv_1}, \eqref{eq:conform_inv_2}, for a conformally invariant critical system, indeed lead to the vanishing of both transport coefficients.

We will now examine the behaviour of the superfluid bulk viscosities \eqref{eq:bulk_visc_conv} close to the superfluid phase transition. In order to take the limit near criticality, we reiterate our assumption that the coefficients $\Gamma_0,\,Z_2,Z_n$  remain finite and that the susceptibilities scale with $\ves$ as outlined in Appendix \ref{app:mean_exp}. It is easy then to see that all three bulk viscosities diverge as $\sim \frac{1}{\ves^2}$. This divergence is a direct manifestation of the breakdown of conventional superfluid hydrodynamics in the nearly critical region. As one would expect, this divergence shows up because we have integrated out a nearly massless degree of freedom.

The quasinormal modes following from \eqref{eq:conv_super}, \eqref{eq:Jos_rel}, upon setting the external sources to zero and solving the equations of motion, are then the five modes of a superfluid: Two first and two second sound modes and a shear momentum, diffusive mode. The behaviour of these modes close to the critical point is described in more detail in \cite{Donos:2022www}. Here we shall simply observe that the attenuation constant for the first sound mode diverges close to the critical point as $\zeta_1\sim\frac{1}{\ves^2}$, but the attenuation constant for the second sound is finite and proportional to $\chi_{JJ}\,\zeta_i\sim \ves^0$.

In the small wavevector limit, apart from conventional superfluid dynamics, our effective theory captures one additional non-hydrodynamic mode, the so-called amplitude/Higgs mode \cite{Donos:2022xfd,Donos:2022qao}. To find its dispersion relation, we expand its frequency according to,
\begin{align}
\omega=\omega_0+\omega_1\,\lambda+\omega_2\,\lambda^2+\cdots\,.
\end{align}
In contrast to \eqref{eq:lambda_scaling_1}, the presence of the $\omega_0$ is needed since the Higgs mode is by definition gapped. From the equations of motion it then follows that $\delta v_i=\mathcal{O}(\lambda^2),\,\delta\theta=\mathcal{O}(\lambda)$ and the dispersion relation for the Higgs mode reads,
\begin{align}\label{eq:dispers_Higgs}
    \omega_{H}=\omega_g-i D_H k^2+\mathcal{O}(k^4)\,,
\end{align}
where the gap is given by
\begin{align}\label{eq:Gap_new}
    \omega_g=-i\frac{2\,\mathrm{Re}\Gamma_0+\mathrm{Re}Z_n\, \rho_v^2}{2\,\tilde{\nu}_{\rho\rho}+\mathrm{Re}Z_\pi}+\mathcal{O}(\ves^6)
\end{align}
and the diffusion constant is
\begin{align}\label{eq:D_higgs}
    D_H=-\frac{(s_c\, \nu_{\rho s}+\varrho_c\,\nu_{\rho \varrho})^2}{(s_c\, T_c+\mu\,\varrho_c)\mathrm{Re}\Gamma_0}+\mathcal{O}(\ves^0)\,.
\end{align}

Let us first comment on this formula for $\omega_g$. Assuming all transport coefficients are finite close to the transition (behave as $\sim \ves^0$ at leading order), then the gap of the Higgs mode to leading order is simply $-i \frac{\mathrm{Re}{\Gamma_0}}{\tilde{\nu}_{\rho\rho}}+\mathcal{O}(\ves^4)$, a result that agrees with the one reported in \cite{Donos:2022xfd,Donos:2022qao}\footnote{See for instance formulas 5.4-5.5 of \cite{Donos:2022xfd}, noting that there the susceptibility $\tilde{\nu}_{\rho\rho}$ was denoted as $\chi_{\mathcal{O}_\rho \mathcal{O}_\rho}$.}. The $\ves^4$ part of the gap, also captured by \eqref{eq:Gap_new}, is a new result. Notice, though, that we cannot trust \eqref{eq:Gap_new} to even higher orders in $\ves$. To see that, we observe that in the next order we could include a term $Z_N\, \psi^2  \hat{D}_u \mathcal{F}_\psi^\ast$ in $E_{diss}$ of \eqref{eq:corr_constit}, with $\mathrm{Re}Z_N$ finite. Then it is straightforward to see that this new coefficient would also contribute to order $\ves^6$ in the expression for the gap.

We now turn our attention to the diffusion constant $D_H$ for which equation \eqref{eq:D_higgs} shows that it diverges like $\sim\frac{1}{\ves^2}$, close to the critical point. This is in contrast to the results of Model F \cite{Halperin:1974zz} which considers only the coupled sector of the order parameter and the charge density. In \cite{Donos:2022qao} we found that, in that system, the same diffusion constant remains finite as $\ves\to 0$. We thus conclude that the finiteness of $D_H$ in that case was just an artefact of the probe limit and the diverging behaviour is the correct one\footnote{ As a cross-check,  we can consider the probe limit in our calculation, taking $T_c\to\infty,\,\nu_{T\rho}\,T_c=\mathrm{finite},\xi\,T_c=\mathrm{finite}$. Doing so, we can see that, indeed, the $\sim \frac{1}{\ves^2}$ part of $D_H$ vanishes, and its finite part agrees with formula 5.9 of \cite{Donos:2022qao} for $D_H$.}.

\subsection{Normal fluid hydrodynamics}\label{sec:Normal_fluid}

We will now examine the limit where the wavevector of our fluctuations is much larger than the gap of the order parameter amplitude mode. In other words, we will take $\ves\to 0$, while keeping $k$ and $\omega$ fixed in the linearised theory of Section \ref{sec:Eff_linear}. More concretely, we consider the scaling\footnote{We again set the complex scalar sources to zero. Also, note that the scaling of $\delta\rho_v$ is dictated by the equation of motion of the order parameter.},
\begin{align}
    \omega,\,k_i=\mathcal{O}(\ves^0),\qquad\delta\tilde{\rho_v}=\frac{\delta\rho_v}{\rho_v}=\mathcal{O}(\ves^0)\,,
\end{align}
with all other perturbations finite in the small $\ves$ limit.

We now take the limit $\ves\to 0$ in \eqref{eq:const_Linear}, \eqref{eq:ord_linear} and use equations \eqref{eq:NP_1}, \eqref{eq:NP_2}, to relate the leading parts of the broken and normal phase susceptibilities close to the critical point to find,
\begin{align}\label{eq:const_fin_normal}
    \delta T^{tt}&=\left(c_\mu^\# +\mu_0\,\xi^\#\right) \delta T+\left(\mu_0\, \chi^\#+T_0\, \xi^\#\right)\delta\mu-\epsilon_0\, \delta g^{tt}\,,\nn
    \delta T^{ti}&=(\epsilon_0+p_0)\delta v^i-\epsilon_0\, \delta g^{ti}\,,\nn
    \delta T^{ij}&=p_0\, \delta g^{ij}+\left(s_0\,\delta T+\varrho_0\,\delta\mu\right)\delta^{ij}- Z_1\, \delta^{ij}\left(\partial_k(\delta v^k-\delta g^{kt})+\partial_t \delta g_{kl}\,\frac{\delta^{kl}}{2}\right)-\eta\,\delta \sigma^{ij}\,,\nn
    \delta J^t&=\xi^\#\, \delta T+\chi^\#\,\delta\mu-\varrho_0\, \frac{\delta g^{tt}}{2}\,,\nn 
    \delta J^i&=\varrho_0\,\delta v^i-\varrho_0\,\delta g^{ti}-\sigma\,\delta^{ij}\left(\partial_j \delta\mu- \frac{\mu_0}{T_0}\,\partial_j\delta T-\delta F_{jt}\right)\,,
\end{align}
after discarding terms of order $\mathcal{O}(\ves^2)$. It is important to note that neither the amplitude nor the phase of the order parameter appear at this (leading) order\footnote{They will first appear at order $\ves^2$.}. In fact, the formulas \eqref{eq:const_fin_normal} are precisely the linearised constitutive relations for the stress tensor and current of a charged normal fluid in the Landau frame, up to first order in the derivative expansion (see e.g. \cite{Kovtun:2012rj}). 

Similarly, dividing \eqref{eq:ord_linear} by $\rho_v$ and taking $\ves$ small, we find the complex equation of motion,
\begin{align}\label{eq:order_eq_nc}
    &\partial_t \delta \tilde{\rho}_v+i q_e \delta m_t+iq_e\,\mu_0\,\frac{\delta g_{tt}}{2}-iq_e\, \delta \mu=\nn& -2\, \overline{\Gamma}_0\,\left( -\frac{w_0}{2}\left(\partial_i^2\delta\tilde{\rho}_v+i q_e \partial_i(\delta m^i+\mu_0\, \delta v^i)\right)-\frac{1}{2\, \nu_{\rho\rho}\, \rho_v}\left(\nu_{T \rho}\,\delta T+\nu_{\mu\rho}\,\delta\mu\right)\right)\nn&
+Z_2\,\left(\partial_k(\delta v^k-\delta g^{kt})+\partial_t \delta g_{kl}\,\frac{\delta^{kl}}{2}\right)\nn&-Z_\pi\, \partial_t\,\left( -\frac{w_0}{2}\left(\partial_i^2\delta\tilde{\rho}_v+i q_e \partial_i(\delta m^i+\mu_0\, \delta v^i)\right)-\frac{1}{2\, \nu_{\rho\rho}\, \rho_v}\left(\nu_{T \rho}\,\delta T+\nu_{\mu\rho}\,\delta\mu\right)\right)\,,
\end{align}
where once again, we have ignored terms of order $\mathcal{O}(\ves^2)$. We also highlight that the hydrodynamic parameters of the normal phase, $\delta T,\,\delta\mu,\,\delta v^i$, do show up in the equation for the order parameter \eqref{eq:order_eq_nc} at this order in $\ves$.
 
Let us now briefly discuss the quasinormal modes in this asymptotic regime, setting the external metric $\delta g_{\mu\nu}$ and gauge field $\delta A_\mu$ to zero. We have two sets of quasinormal modes; The first set consists of four quasinormal modes following from \eqref{eq:const_fin_normal}, which are naturally interpretable as the quasinormal modes of a charged normal fluid\cite{Kovtun:2012rj}: Two sound modes, a shear momentum diffusion mode, and a charge diffusion mode. In these modes, though, the amplitude and the phase of the order parameter are not trivial, since $\delta T,\,\delta\mu,\,\delta v^i$ act like ``source terms" in \eqref{eq:order_eq_nc}. The second set consists of the  quasinormal modes following from \eqref{eq:order_eq_nc}. For this set of modes the stress tensor and current conservation equations dictate that the trivial solution  $\delta T=\delta\mu=\delta v^i=0$ is the only one, since the determinant of coefficients of the corresponding 3 by 3 linear system is not zero.

 Hence, for the second set of quasinormal modes, \eqref{eq:order_eq_nc} further simplifies to
\begin{gather}\label{eq:nc_diff}
\partial_t \delta \tilde{\rho}_v+i q_e \partial_t \delta\theta= \left(\overline{\Gamma}_0+\frac{Z_\pi}{2}\partial_t\right)w_0\left(\partial_i^2\delta\tilde{\rho}_v+iq_e\, \partial_i^2\delta\theta \right).
\end{gather}

This gives two diffusive modes,
\begin{align}\label{eq:complex_normal}
    \omega_{r,1}=-iw_0\,\Gamma_0\,k^2+\mathcal{O}(k^4),\quad\omega_{r,2}=-iw_0\,\overline{\Gamma}_0\,k^2+\mathcal{O}(k^4)\,.
\end{align}

As shown in \cite{Donos:2022qao}, these modes match precisely with the quasinormal modes of the order parameter in the normal phase, close to the critical point. (Notice that we cannot trust the $O(k^4)$ part of the modes following from \eqref{eq:nc_diff}, in which $Z_\pi$ appears. The reason is that at the same order there are additional terms that would contribute, e.g. $(D_\mu^\perp)^2 \mathcal{F}_\psi$. These were not considered in \eqref{eq:corr_constit}, due to the relative scaling $\omega\sim k$ assumed there.)

\section{Keldysh-Schwinger effective theory
}\label{sec:KS_hydro}

In this section, we present a Keldysh-Schwinger \cite{Crossley:2015evo, Haehl:2015uoc, Liu:2018kfw} construction of the effective theory for nearly critical superfluids. We will assume that in the regime of the critical point, thermal fluctuations are dominant over quantum effects. This will allow us to work in the pure classical limit ($\hbar\to0$) and make use of the physical spacetime formulation \cite{Crossley:2015evo}, \cite{Glorioso:2017fpd}. We will build on our previous work \cite{Donos:2023ibv}, essentially generalising it beyond the probe limit\footnote{See also \cite{Kapustin:2022iih}, \cite{Bu:2024oyz} for work on the same topic.}, so that we also include temperature and normal fluid velocity fluctuations. In our notation and general discussion of the formalism, we closely follow \cite{Glorioso:2017fpd}.

\subsection{Setup}\label{sec:Setup_KS}
In the context of this formalism, the natural variables needed to describe a charged normal fluid are the inverse temperature $\beta$, the normal fluid velocity $u^\mu$ and the dynamical variable $\phi$ for the charge density of the system. The dynamics of the superfluid component of the system will be captured by the complex order parameter $\psi$ with external source $s_\psi$. We assume that our fluid lives in a physical spacetime with coordinates $X^\mu$, metric $g_{\mu\nu}$ and $U(1)$ gauge field $A_{\mu}$. After introducing  the Lagrangian fluid element coordinates $\sigma^A$ and the corresponding induced metric $h_{AB}$, the normalised fluid velocity is $u^\mu=\frac{1}{\sqrt{-h_{00}}}\frac{\partial X^\mu}{\partial \sigma^0}$.  The above constitute the $r$-fields of our system. The corresponding $a$-fields are the metric $G_{a\mu\nu}$, the gauge field $A_{a\mu}$, the phase $\phi_a$ and the order parameter $\psi_a$ with source $s_{\psi_a}$.

One of the main goals of this section is to construct an effective action $I_{EFT}=\int d^{d}x \sqrt{-g}\,\mathcal{L}_{EFT}$, as a functional of our degrees of freedom and sources. The first requirement that our effective action has to satisfy is the well known unitarity conditions,
\begin{align}\label{eq:unitarity}
&I_{EFT}[\Lambda_r,\,\Lambda_a=0]=0\,,\nn
    &I_{EFT}[\Lambda_r,\,-\Lambda_a]^\ast=-I_{EFT}[\Lambda_r,\,\Lambda_a]\,,\nn
    &\mathrm{Im}\left(I_{EFT}[\Lambda_r,\,\Lambda_a]\right)\geq 0\,,
\end{align}
where $\Lambda_r,\,\Lambda_a$ collectively denote the fields and external sources of $r$ and $a$ type. This is a generic requirement that all effective theories have to satisfy.

For the system we want to examine in our case, the effective action must be separately invariant under diagonal and anti-diagonal $U(1)$ gauge transformations, parametrised by  $\lambda_D$ and $\lambda_A$ respectively. The fields charged under these symmetries transform according to,
\begin{align}
\phi'&=\phi+\lambda_D,\quad\phi'_a=\phi_a+\lambda_A,\nn
    A'_\mu&=A_\mu-\partial_\mu\lambda_D,\quad A_{a\mu}'=A_{a\mu}-\partial_\mu\lambda_A+\nabla_\mu(\mathcal{L}_{X_a}\lambda_D),\nn
\psi'&=e^{iq_e\lambda_D}\psi,\quad\psi'_a=e^{iq_e\lambda_D}(\psi_a+i\,q_e\,\lambda_A\,\psi)\,,\nn
s_\psi'&=e^{iq_e\lambda_D}s_\psi,\quad s'_{\psi_a}=e^{iq_e\lambda_D}(s_{\psi_a}+i\,q_e\,\lambda_A\,s_\psi)\,.
\end{align}
In order to construct our effective action, we found it convenient to introduce the alternative variables,
\begin{align}\label{eq:gauge_inv_SK}
    B_\mu&=\partial_\mu\phi+A_\mu\,,\quad C_{a\mu}=\partial_\mu \phi_a+A_{{a\mu}}+\mathcal{L}_{X_a}A_\mu\,,\nn
    \hat{\psi}&=e^{-iq_e\phi}\psi\,,\quad \hat{\psi}_a=e^{-iq_e\phi}\left(\psi_a-iq_e\,\phi_a\,\psi\right)\,,\nn
    \hat{s}_\psi&=e^{-iq_e\phi} s_\psi\,,\quad \hat{s}_{\psi_a}=e^{-iq_e\phi}\left(s_{\psi_a}-iq_e\,\phi_a\,s_\psi\right)\,,
\end{align}
which are invariant under the $U(1)$ gauge transformations.

In order to correctly describe the dynamics of the local charge density, the theory must also be invariant under a ``chemical shift'' transformation with time independent parameter $\hat{\lambda}(\sigma^i)$ satisfying $\nabla\hat{\lambda}=\nabla^\perp \hat{\lambda}$. This transformation only shifts $\phi\to\phi-\hat{\lambda}$ in the notation of our original fields. However, since the definitions \eqref{eq:gauge_inv_SK} of our gauge invariant variables involve $\phi$, they transform according to,
\begin{align}\label{eq:chem_trans}
    \hat{\psi}\to e^{iq_e\hat{\lambda}}\hat{\psi}\,,\quad \hat{\psi}_a\to e^{iq_e\hat{\lambda}}\hat{\psi}_a\,,\quad B_\mu\to B_\mu-\nabla^\perp_\mu \hat{\lambda}\,,
\end{align}
under a chemical shift. Notice that the combinations $\mu\equiv u^\mu\, B_\mu$ and $F_{\mu\nu}=\nabla_\mu B_\nu-\nabla_\nu B_\mu$, are invariant under the chemical shift symmetry. We will take the combination $\mu$ as our definition of the out of equilibrium chemical potential. We notice that we can write $B_\mu = B^\perp_\mu - \mu\,u_\mu$, which implies that the longitudinal component of the one form field $B_\mu$ is of order zero in derivatives. However, the projected component $B^\perp_\mu$ has to be considered as first order in our derivative expansion scheme. This becomes clear after noticing that the chemical shift parameter $\hat{\lambda}$ is of order $\mathcal{O}(\partial^0)$. Moreover, its gradient mixes only with $B^\perp_\mu$. We should also remark that the field strength can be written as $F_{\mu\nu}=\nabla_\mu B_\nu^{\perp}-\nabla_\nu B_\mu^\perp-\nabla_\mu(\mu u_\nu)+\nabla_\nu(\mu u_\mu)$. Hence $F_{\mu\nu}$ contains both terms of order $\mathcal{O}(\partial)$ and terms of order $\mathcal{O}(\partial^2)$.

For our purposes, it is natural to introduce a covariant derivative under chemical shifts,
\begin{align}
    D_{\mu}\hat{\psi}=\nabla_\mu \hat{\psi}+iq_e \,B^\perp_\mu\, \hat{\psi}\,,\quad D_{\mu}\hat{\psi}_a=\nabla_\mu \hat{\psi_a}+iq_e \,B^\perp_\mu\, \hat{\psi_a}\,,
\end{align}
where $B^\perp_\mu=P_\mu{} ^\nu B_\nu$ and  $P^{\mu\nu}=g^{\mu\nu}+u^\mu u^\nu$ is the projection operator normal to $u^\mu$. For later convenience, let us also introduce the thermodynamic quantities $\tau\equiv \ln\left( \beta/\beta_0\right)$ for a constant $\beta_0$ and $\hat{\mu}\equiv \beta\,\mu$.

Finally, given an antiunitary discrete symmetry generator $\Theta$, our effective action must be invariant under a corresponding dynamical KMS transformation. In our case, we will take $\Theta=\mathcal{P}\,\mathcal{T}$ to be a combination of parity and time reversal. Given this choice, the dynamical KMS transformation rules read\footnote{These transformations are correct in the \emph{exact} classical limit $\hbar\to 0$ \cite{Liu:2018kfw}. Including $\mathcal{O}(\hbar)$ corrections would bring additional terms, with more $\mathcal{L}_{\beta^\rho}$ derivatives, in the transformation laws.} 
\begin{align}\label{eq:KMS}
    \tilde{u}^\mu(-x)&=u^\mu(x),\quad\tilde{\beta}(-x)=\beta(x),\quad\tilde{B}_\mu(-x)=B_\mu(x),\quad \tilde{g}_{\mu\nu}(-x)=g_{\mu\nu}(x)\,,\nn
    \tilde{\hat{\psi}}(-x)&=\hat{\psi}^\ast(x),\quad\tilde{\hat{\psi}}^\ast(-x)=\hat{\psi}(x),\quad \tilde{\hat{s}}_\psi(-x)=\hat{s}_\psi^\ast(x),\quad\tilde{\hat{s}}_\psi^\ast(-x)=\hat{s}_\psi(x)\,, \nn  
    \tilde{G}_{a\mu\nu}(-x)&=G_{a\mu\nu}(x)+i\,\mathcal{L}_{\beta^\rho}g_{\mu\nu}(x),\quad\tilde{C}_{a\mu}(-x)=C_{a\mu}(x)+i\,\mathcal{L}_{\beta^\rho}B_{\mu}(x),\nn
    \tilde{\hat{\psi}}_a(-x)&=\hat{\psi}^\ast_a(x)+i\,\mathcal{L}_{\beta^\rho}\hat{\psi}^\ast(x),\quad \tilde{\hat{\psi}}_a^\ast(-x)=\hat{\psi}_a(x)+i\,\mathcal{L}_{\beta^\rho}\hat{\psi}(x)\,,\nn
     \tilde{\hat{s}}_{\psi_a}(-x)&=\hat{s}^\ast_{\psi_a}(x)+i\,\mathcal{L}_{\beta^\rho}\hat{s}_\psi^\ast(x),\quad \tilde{\hat{s}}_{\psi_a}^\ast(-x)=\hat{s}_{\psi_a}(x)+i\,\mathcal{L}_{\beta^\rho}\hat{s}_\psi(x)\,,
\end{align}
with $\mathcal{L}_{\beta^\rho}$ being the Lie derivative along $\beta^\mu=\beta\, u^\mu$.

In the Keldysh-Schwinger framework, a natural expansion scheme for $\mathcal{L}_{EFT}$ is in terms of the number of $a$-fields that appear in each term. Schematically, we can think of organising our effective Lagrangian according to, $\mathcal{L}_{EFT}=\mathcal{O}(a)+ \mathcal{O}(a^2)+\cdots$, where we have exploited that the first of the unitarity constraints \eqref{eq:unitarity} requires the absence of $\mathcal{O}(a^0)$ terms. Moreover, the KMS transformation rules \eqref{eq:KMS} (almost) preserve the total number of derivatives and $a$-fields. We say almost since the KMS rule for $C_{a\mu}$ can be written as $\tilde{C}_{a\mu}(-x)=C_{a\mu}(x)+i\beta^\nu F_{\nu \mu}+i \nabla_\mu \hat{\mu}$, and as we mentioned below \eqref{eq:chem_trans}, $F_{\mu\nu}$ contains $\mathcal{O}(\partial)$ terms, but also the term $\nabla_\mu B_\nu^{\perp}-\nabla_\nu B_\mu^\perp$, which within our counting is considered $\mathcal{O}(\partial^2).$ Based on this observation, one expands the effective Lagrangian as
\begin{align}\label{eq:L_expansion}
    \mathcal{L}_{EFT}=\sum_{n=1}^{\infty}\mathcal{L}_{[n]}\,,
\end{align}
where $\mathcal{L}_{[n]}$ includes all terms in which the total number of derivatives and $a$-field factors is $n$. Examining how each term in our series transforms under the KMS transformation rules, we can write\footnote{Based on our observation above, this equation is correct, except for possible terms including $\nabla_\mu B_\nu^{\perp}-\nabla_\nu B_\mu^\perp$ on its right hand side.},
\begin{align}\label{eq:KMS_use}
    \tilde{\mathcal{L}}_{[n]}(-x)=\mathcal{L}_{[n]}(x)+i\,\nabla_\mu V^\mu_{(0,\,n-1)}+\nabla_\mu V^\mu_{(1,\, n-2)}\,,
\end{align}
where in $V^\mu_{(i,j)}$, the index $i$ counts the number of $a$-fields and $j$ the number of derivatives. One would naively expect terms with more $a$-fields to appear on the right hand side. However, as shown in Appendix B of \cite{Glorioso:2017fpd}, by exploiting the $\mathbb{Z}_2$ property of the KMS transformation, these can be absorbed in $\mathcal{L}_{[n]}$ after an integration by parts. Note that $V^\mu_{(1,\,n-1)}$ can be nontrivial, given our assumption that the $\mathcal{O}(a)$ terms of $\mathcal{L}_{EFT}$ do not include derivatives of $a$-fields.

\subsection{Effective action}\label{sec:Eff_action}
The most general Lagrangian term for $n=1$ reads,
\begin{align}\label{eq:L_1}
\mathcal{L}_{[1]}=2\,\mathrm{Re}\left[\hat{\psi}_a^\ast\left(\frac{\partial p_0}{\partial |\hat{\psi}|^2}\hat{\psi}+\frac{\hat{s}_\psi}{2}\right)+\frac{\hat{s}_{\psi_a}^\ast}{2}\,\hat{\psi}\right]+C_{a\mu}\,J^\mu_0+\frac{G_{a\mu\nu}}{2}T^{\mu\nu}_0\,,
\end{align}
with
\begin{align}
    J^\mu_0=\varrho_0\, u^\mu,\quad T^{\mu\nu}_0=\left(\epsilon_0-\mathrm{Re}\left(\hat{s}_\psi^\ast\,\hat{\psi}\right)\right)\, u^\mu u^\nu+\left(p_0+\mathrm{Re}\left(\hat{s}_\psi^\ast\,\hat{\psi}\right)\right)\, P^{\mu\nu}\,
\end{align}
and $\epsilon_0,\, p_0,\,\varrho_0$ arbitrary real functions of $\mu, \beta, |\hat{\psi}|^2$ obeying the KMS constraints,
\begin{align}
  \frac{\partial  p_0}{\partial \mu} =\varrho_0\,,\quad   \frac{\partial  p_0}{\partial \beta}=-\frac{p_0+\epsilon-\mu\, \varrho_0}{\beta}\,.
\end{align}
These are precisely the relation between the pressure $p_0$ and the charge density along with the thermodynamic relation between entropy, pressure, charge and energy. By performing a KMS transformation on \eqref{eq:L_1}, we can easily check that the condition \eqref{eq:KMS_use} is satisfied with $V^\mu_{(0,\,0)}= \left(p_0+\mathrm{Re}\left(\hat{s}_\psi^\ast\,\hat{\psi}\right)\right)\, \beta^\mu$. Moreover, the above results for $n=1$ are a direct extension of the $n=1$ action of \cite{Glorioso:2017fpd}, with the inclusion of the condensate\footnote{See in particular Section V therein.}.

For the $n=2$ term in the Lagrangian expansion \eqref{eq:L_expansion}, the most general expression, compatible with the first two unitarity constraints \eqref{eq:unitarity} reads,
\begin{align}\label{eq:L_2}
\mathcal{L}_{[2]}&=2\,\mathrm{Re}\left(\hat{\psi}^\ast_a\, E_1\right)+C_{a\mu}\, J^\mu_1+\frac{G_{a\mu\nu}}{2}T_1^{\mu\nu}+\frac{i}{4}W_0^{\mu\nu,M N}G_{a\mu M}G_{a\nu N}+\nn & +2i\,\mathrm{Im}\left[\lambda_0^{\mu M}G_{a \mu M}\,\hat{\psi}_a\,\hat{\psi}^\ast+\kappa_{1}\hat{\psi}_a^2 \left(\hat{\psi}^\ast\right)^2\right]+i\, \kappa_{0}|\hat{\psi}_a|^2\,,
\end{align}
where we have used the notation $G_{a\mu N}=(G_{a\mu\nu},2\,C_{a\mu})$ of \cite{Glorioso:2017fpd}. The quantity $W_0^{\mu\nu,M N}$ is a real tensor structure without derivatives, which we list in Appendix \ref{app:nf_corrections} and which we can directly import from \cite{Glorioso:2017fpd}. The only difference is that the coefficients $s_{11},s_{22}$ etc. appearing in it
can now also depend on $|\hat{\psi}|^2$. The coefficients $\kappa_0, \kappa_1$ are arbitrary scalar functions of $\mu, \beta, |\hat{\psi}|^2$ and $\kappa_0$ is real, while the coefficients $\lambda_0^{\mu N}=(\lambda^{\mu\nu}_0, \lambda^\mu_0)$ are given by
\begin{align}
    \lambda^{\mu\nu}_0=\lambda_{01}\,u^\mu u^\nu+\lambda_{02}\, P^{\mu\nu},\quad \lambda^\mu_0=\lambda_{03}\,u^\mu\,,
\end{align}
 with $\lambda_{0 i}$ complex functions of $\mu,\beta, |\hat{\psi}|^2$. 
 
The first order derivative corrections of the stress tensor and current can be decomposed to a normal fluid and a condensate part according to $T_1^{\mu\nu}=T_{1, n}^{\mu\nu}+T_{1, \psi}^{\mu\nu}$, $J^\mu_1=J^\mu_{1,n}+J^\mu_{1,\psi}$. The second terms in this decomposition contain only corrections involving derivatives of the order parameter $\hat{\psi}$. In Appendix \ref{app:nf_corrections}, we have listed the normal fluid corrections $T_{1, n}^{\mu\nu}$ and $J^\mu_{1,n}$ which can be directly read off from e.g. \cite{Glorioso:2017fpd}. For the order parameter derivative corrections we can write the most general expressions,
 \begin{align}\label{eq:SK_stress_1}
     T^{\mu\nu}_{1,\psi}&=\epsilon_\psi\, u^{\mu}u^\nu+p_\psi\, P^{\mu\nu}+2\, u^{(\mu}q^{\nu)}_\psi ,\quad J^\mu_{1,\psi}=\varrho_\psi\, u^\mu+j^\mu_\psi\,,\nn
     \epsilon_\psi &= \mathrm{Re}\left(d_{21}\, \hat{\psi}^\ast u^\mu D_\mu \hat{\psi}\right),\quad p_\psi= \mathrm{Re}\left(d_{22}\, \hat{\psi}^\ast u^\mu D_\mu \hat{\psi}\right),\quad \varrho_\psi= \mathrm{Re}\left(d_{11}\, \hat{\psi}^\ast u^\mu D_\mu \hat{\psi}\right),\nn
     q^\mu_\psi&=\mathrm{Re}\left(c_2\, \hat{\psi}^\ast D^{\perp\mu}\hat{\psi} \right),\quad j^\mu_\psi=\mathrm{Re}\left(c_1\, \hat{\psi}^\ast D^{\perp\mu}\hat{\psi} \right)\,.
 \end{align}

Finally, $E_1$ which appears in \eqref{eq:L_2} is a scalar quantity which also contains first order derivatives at most and it can be seen as a dissipative correction to the classical equation of motion of the order parameter. More concretely, the most general expression takes the form,
\begin{align}\label{eq:SK_eq_1}
    E_1=c_{11}\,u^\mu D_\mu \hat{\psi}+c_{15}\,\hat{\psi}^2\,u^\mu D_\mu \hat{\psi}^\ast+\hat{\psi}\left(c_{12}\,\nabla_\mu u^\mu+c_{13}\,u^\mu\nabla_\mu\, \beta+c_{14}\,u^\mu\nabla_\mu\,\mu\right)\,.
\end{align}
At this point, we should note that all coefficients appearing in \eqref{eq:SK_stress_1} and \eqref{eq:SK_eq_1} are complex functions of $\mu, \beta, |\hat{\psi}|^2$.

This concludes the most general form of the $n=2$ in our Lagrangian expansion \eqref{eq:L_expansion} in terms of a set of transport coefficients. The final ingredient in our construction is to impose that the specific term satisfies equations \eqref{eq:KMS_use} under the dynamical KMS transformation rules \eqref{eq:KMS}. This requirement yields the following constraints on the new coefficients we have introduced,
\begin{align}\label{eq:KMS_2}
    &\lambda_{01}=i\,\mathrm{Re}(c_{13})-i\frac{\mu}{\beta}\mathrm{Re}(c_{14}),\quad\lambda_{02}=-\frac{i}{\beta}\mathrm{Re}(c_{12}),\quad\lambda_{03}=-\frac{i}{\beta}\mathrm{Re}(c_{14}),\nn
    &\kappa_0=-\frac{2}{\beta}\mathrm{Re}(c_{11}),\quad \kappa_1=-\frac{i}{\beta}\mathrm{Re}(c_{15})\,,\nn
&d_{11}=2\,c_{14},\quad d_{22}=2\,c_{12},\quad d_{21}=-2\,\beta\,c_{13}+2\mu\, c_{14},\nn
    &\mathrm{Im}(c_{15})=\mathrm{Re}(c_1)=\mathrm{Re}(c_2)=0,\quad \mathrm{Im}(c_2)=\mu\,\mathrm{Im}(c_1)\,.
\end{align}
The coefficients appearing in $W_0^{\mu\nu,MN}$, $T_{1, n}^{\mu\nu}$, and $J^\mu_{1,n}$ obey exactly the same constraints as those found in \cite{Glorioso:2017fpd} and we list them in Appendix \ref{app:nf_corrections} for completeness.

After taking into account the above constraints, we are left with the KMS transformation rule,
\begin{align}\label{eq:KMS_change_2}
    \tilde{\mathcal{L}}_{[2]}(-x)=\mathcal{L}_{[2]}(x)-i\beta^\rho(\nabla_\rho B_\mu^\perp-\nabla_\mu B_\rho^\perp)\,\mathrm{Im}\left(\hat{\psi}^\ast D^{\perp\mu}\hat{\psi}\right)\mathrm{Im}(c_1)\,,
\end{align}
for our $n=2$ term, which is clearly not a total divergence. However, as we discussed earlier, the projected component $B_\mu^\perp$ has to be considered as first order in the expansion. This turns the non-divergence term of the transformation \eqref{eq:KMS_change_2} to be third order in derivatives. Therefore, such a term has to be cancelled by terms in $\mathcal{L}_{[3]}$, as we will show below.

For $n=3$, the Lagrangian is much more involved\footnote{As far as we know, even the effective Lagrangian for a charged normal fluid has not been written in full generality.}. Here we will be interested in finding a truncated version $\mathcal{L}_{[3],tr.}$ of the most general $\mathcal{L}_{[3]}$, which is KMS invariant and includes second derivatives of the condensate. A sufficiently general ansatz is,
\begin{align}
    \mathcal{L}_{[3],tr.}=\frac{G_{a\mu\nu}}{2}T^{\mu\nu}_2+C_{a\mu}\,J^\mu_2+2\,\mathrm{Re}\left(\hat{\psi}^\ast_a\, E_2\right)+2 i\, \mathrm{Im}\left(e_{33}\,\hat{\psi}^\ast_a\, \beta^\mu D_\mu \hat{\psi}_a\right)\,,
\end{align}
with,
\begin{align}
    T_2^{\mu\nu}&=e_{11}\,D^{\perp(\mu}\hat{\psi} D^{\perp\nu)}\hat{\psi}^\ast+\left(e_{12}\,u^\mu u^\nu+e_{13}\,P^{\mu\nu}\,\right)|D_\rho^\perp \hat{\psi}|^2,\nn
    J_2^\mu&=e_{14}|D_\rho^\perp \hat{\psi}|^2\, u^\mu,\nn
    E_2&=\frac{1}{2\beta}D_\mu^\perp\left(e_{21}\,\beta\,D^{\perp \mu}\hat{\psi}\right)+e_{22}\,\left(\beta^\mu D_\mu\right)^2\hat{\psi}+\nabla_\mu\left( e_{23}\,\beta^\mu\right)\beta^\nu D_\nu\hat{\psi}.
\end{align}
The action term $\mathcal{L}_{[3],tr.}$ obeys the first two constraints \eqref{eq:unitarity} by construction, provided that the coefficients $e_{1i}$ are real.

Demanding KMS invariance implies that our transport coefficients have to satisfy the relations,
\begin{align}
    e_{23}&=\frac{e_{22}}{2},\quad e_{33}=\mathrm{Im}(e_{22}),\quad  e_{21}=-2\,e_{13}=e_{11}\,,\quad \mathrm{Im}(c_1)=q_e\, e_{11}\,,\nn
    \frac{\partial e_{11}}{\partial \mu}&=-2\, e_{14}\,,\quad \frac{\partial e_{11}}{\partial \beta}=\frac{2}{\beta}\left(e_{12}-\mu\, e_{14}-\frac{e_{11}}{2}\right)\,,\quad  \frac{\partial e_{11}}{\partial \hat{\psi}}=0\,.
\end{align}
Given the above constraints, our $n=3$ truncated Lagrangian term transforms according to,
\begin{align}\label{eq:KMS_change_3}
    \tilde{\mathcal{L}}_{[3],tr.}(-x)=\mathcal{L}_{[3], tr.}(x)&+i \nabla_\mu V^\mu_{(0,2)}+\nabla_\mu V^\mu_{(1,1)}+\nn
    &+iq_e\,e_{11}\beta^\rho(\nabla_\rho B_\mu^\perp-\nabla_\mu B_\rho^\perp)\,\mathrm{Im}\left(\hat{\psi}^\ast D^{\perp\mu}\hat{\psi}\right)\,,
\end{align}
with,
\begin{align}
    V^\mu_{(0,2)}&=-\frac{e_{11}}{2}\beta^\mu|D_\rho^\perp \hat{\psi}|^2+e_{11}\,\mathrm{Re} \left[D^{\perp\mu}\hat{\psi}^\ast \beta^\nu D_\nu \hat{\psi}\right]+\beta^\mu\,\mathrm{Re}
    (e_{22})|\beta^\nu D_\nu \hat{\psi}|^2\,,\nn
   V^\mu_{(1,1)}&=2\,\mathrm{Im}(e_{22})\,\beta^\mu\, \mathrm{Im}\left[\hat{\psi}_a^\ast\,\beta^\nu D_\nu \hat{\psi} \right]\,.
\end{align}
We note that the last term of \eqref{eq:KMS_change_3} indeed cancels out with the last term of \eqref{eq:KMS_change_2}.

Following \cite{Glorioso:2016gsa}, the entropy current for our theory is defined as,
\begin{align}\label{eq:entr_SK}
    s^\mu=V_0^\mu-\hat{V}_1^\mu-T^{\mu\nu}\beta_\nu-J^\mu \hat{\mu}
\end{align}
with,
\begin{align}
    V^\mu_0&=V^\mu_{(0,0)}+V^\mu_{(0,2)}=p\, \beta^\mu+e_{11}\,\mathrm{Re} \left[D^{\perp\mu}\hat{\psi}^\ast \beta^\nu D_\nu \hat{\psi}\right]+\beta^\mu\,\mathrm{Re}
    (e_{22})|\beta^\nu D_\nu \hat{\psi}|^2\,,\nn
    \hat{V}^\mu_1&=2\,\mathrm{Im}(e_{22})\,\beta^\mu\, \mathrm{Im}\left[\beta^\rho D_\rho\hat{\psi}^\ast\,\beta^\nu D_\nu \hat{\psi} \right]=0\,.
\end{align}
In the above relation, we introduce the equilibrium pressure $p$, which includes the gradient correction and the external source of the order parameter,
\begin{align}
p=p_0+\mathrm{Re}\left(\hat{s}_\psi^\ast\,\hat{\psi}\right)-\frac{e_{11}}{2}|D_\rho^\perp\hat{\psi}|^2\,,
\end{align}
in accordance with \eqref{eq:f_tot_def} and \eqref{eq:therm_def}, upon identifying $w_0$ with $e_{11}$.

Closing this subsection, we should again stress that our analysis for $n=1$ and $n=2$ is complete. However,  $\mathcal{L}_{[3].tr}$ is only a truncated version of the most general Lagrangian $\mathcal{L}_{[3]}$, keeping  up to two derivatives of the condensate and which makes the total Lagrangian KMS invariant. This means that the entropy current given in \eqref{eq:entr_SK} is complete only up to first derivative terms; there are additional contributions with two derivatives of $r-$fields, which we have dropped\footnote{All these additional terms and the $\sim\mathrm{Re}
(e_{22})$ term of $s^\mu$ are part of $\Delta s^\mu$ in \eqref{eq:entr_current}.}.

\subsection{Mean field limit}

In this last subsection, we will consider the mean field limit of the theory constructed in \ref{sec:Eff_action}. The path integral over the $r$ and $a$-fields will have to be evaluated in a saddle point approximation, with all the dynamical fields placed on-shell. In this limit, all statistical fluctuations\footnote{Recall that we work in the $\hbar\to0$ limit, where all the \emph{quantum} fluctuations are ignored from the start.} are suppressed, and we recover classical hydrodynamics\cite{Liu:2018kfw}.

Since all the terms in the effective action $I_{EFT}$ contain at least one $a$-field,  the $r$-field equations of motion (with all $a$-field sources vanishing) simply amount to setting all the $a$-fields to zero. On the other hand, the equations of motion for $X^\mu_a$ and $\phi_a$ are the continuity equations \eqref{eq:contin} \cite{Glorioso:2017fpd}. To see that, one must use the invariance of $I_{EFT}$ under $a$-field diffeomorphisms and anti-diagonal gauge transformations, and recall that the on-shell stress tensor and current are given by,
\begin{align}\label{SK_const}
    T^{\mu\nu}&=\frac{2}{\sqrt{-g}}\frac{\delta I_{EFT}}{\delta g_{a\mu\nu}}=T_0^{\mu\nu}+T_1^{\mu\nu}+T^{\mu\nu}_2+\cdots,\nn J^\mu&=\frac{1}{\sqrt{-g}}\frac{\delta I_{EFT}}{\delta A_{a\mu}}=J^\mu_0+J^\mu_1+J^\mu_2+\cdots\,.
\end{align}
Finally, the variation with respect to $\hat{\psi}^\ast_a$ leads to the order parameter equation of motion,
\begin{align}\label{SK_eom}
    \frac{\partial p_0}{\partial |\hat{\psi}|^2}\hat{\psi}+\frac{\hat{s}_\psi}{2}+E_1+E_2+\cdots=0\,.
\end{align}

Our goal now is to compare the constitutive relations \eqref{SK_const} and the order parameter equation \eqref{SK_eom}, following from $I_{EFT}$, with the hydrodynamic theory of Section \ref{sec:Eff_th}. We note that the stress tensor and current in \eqref{SK_const} are written in an arbitrary fluid frame, whereas the effective theory of Section \ref{sec:Eff_th} is written in the transverse frame \eqref{eq:trans_frame_cond}.  We will thus have to perform a field redefinition of $u^\mu, \mu,\,\beta$, in order to fix our hydrodynamic variables to the transverse frame. Furthermore, it will be convenient to express the hatted variables $\hat{\psi},\,\hat{s}_\psi$ in terms of the unhatted ones $\psi,\,s_\psi$ using \eqref{eq:gauge_inv_SK}. In addition, we will have to use the leading order continuity equations to write our expressions in terms of on-shell independent data. The specific relations are as follows,
\begin{align}\label{eq:lead_sol}
    \partial_u \tau&=\partial_\epsilon p_0\,\nabla_\mu u^\mu+c_{\tau\psi}\,\partial_u|\hat{\psi}|^2+c_{\tau s}\,\mathrm{Im}(\hat{\psi}^\ast \hat{s}_\psi)\,,\nn
    \partial_u \hat{\mu}&=-\beta\, \partial_{\varrho}p_0\, \nabla_\mu u^\mu+\beta\,c_{\mu\psi}\,\partial_u|\hat{\psi}|^2+\beta\,c_{\mu s}\,\mathrm{Im}(\hat{\psi}^\ast \hat{s}_\psi)\,,\nn
    \partial_u u^\mu&=P^{\mu\nu}\partial_\nu \tau-\frac{\varrho_0}{\epsilon_0+p_0}\left(P^{\mu\nu}\frac{\partial_\nu \hat{\mu}}{\beta}+u_\nu F^{\nu\mu}\right)\,,
\end{align}
with\footnote{The derivatives of the energy and charge density are calculated in the $\tau,\,\hat{\mu},|\psi|^2$ ensemble and the derivatives of the pressure in \eqref{eq:lead_sol} are calculated in the $\epsilon_0,\varrho_0, |\psi|^2$ ensemble.} 
\begin{align}
    c_{\tau\psi}&=\frac{1}{M}\left(-\partial_{|\psi|^2}\varrho_0\,\partial_{\hat{\mu}}\epsilon_0+\partial_{|\psi|^2}\epsilon_0\,\partial_{\hat{\mu}}\varrho_0\right),\quad c_{\tau s}=\frac{q_e}{M}\left(\partial_{\hat{\mu}}\epsilon_0-\mu\,\partial_{\hat{\mu}}\varrho_0\right)\,,\nn
    c_{\mu\psi}&=\frac{1}{\beta\,M}\left(-\partial_{|\psi|^2}\epsilon_0\,\partial_{\tau}\varrho_0+\partial_{|\psi|^2}\varrho_0\,\partial_{\tau}\epsilon_0\right)\,,\quad c_{\mu s}=\frac{q_e}{\beta\,M}\left(-\partial_{\tau}\epsilon_0+\mu\,\partial_{\tau}\varrho_0\right)\,,
\end{align}
and $M=\partial_{\hat\mu}\epsilon_0\,\partial_\tau \varrho_0-\partial_\tau \epsilon_0\, \partial_{\hat\mu}\varrho_0\,.$

At this point, it is important to observe that the effective action $I_{EFT}$ was constructed solely in a derivative (or $\lambda$) expansion, without the need for an extra $\ves$  expansion. This is naturally introduced when going on-shell, since \eqref{SK_eom} expresses $\frac{\partial p_0}{\partial |\psi|^2}\sim\ves^2$ in terms of derivatives of the effective theory variables. It is straightforward to check that in the scaling region $\lambda\sim \ves^2$, after the necessary change of frame we mentioned, the constitutive relations \eqref{SK_const} and the order parameter equation \eqref{SK_eom} take precisely the form given in \eqref{eq:con_rel_therm},  \eqref{eq:corr_constit}. The coefficients of the hydrodynamic theory, in terms of the coefficients appearing in $I_{EFT}$ read,
\begingroup
\allowdisplaybreaks
\begin{align}\label{eq:Rel_hydro_Sk} 
\Gamma_0&=-\frac{1}{2\,c_{11}^\ast}+\frac{|\psi|^2}{2}\left(c_{\mu s}^2f_{33}+2c_{\tau s}c_{\mu s} f_{13}+c_{\tau s}^2 f_{11}\right)-\frac{|\psi|^2}{2\left(c_{11}^\ast\right)^2}\left(c_{\mu\psi}^2f_{33}+2c_{\tau\psi}c_{\mu\psi}f_{13}+c_{\tau\psi}^2f_{11}\right)\nn&+\frac{i|\psi|^2}{c_{11}^\ast}\left(c_{\tau s}c_{\tau \psi}f_{11}+(c_{\tau\psi}c_{\mu s}+c_{\tau s}c_{\mu \psi})f_{13}+c_{\mu s}c_{\mu\psi}f_{33}\right)\nn&+\frac{|\psi|^2}{2 \left(c_{11}^\ast\right)^2}\left(\left(c_{\mu\psi}-i c_{\mu s}c_{11}^\ast\right)d_{11}^\ast-\left(c_{\tau\psi}-i c_{\tau s}c_{11}^\ast\right)d_{21}^\ast\right)+\mathcal{O}(|\psi|^4)\,, \nn
    Z_n&=\frac{c_{15}}{|c_{11}|^2}-\left(c_{\tau s}^2f_{11}+2c_{\tau s}c_{\mu s}f_{13}+c_{\mu s}^2f_{33}\right)-\frac{1}{|c_{11}|^2}\left(c_{\tau \psi}^2f_{11}+2c_{\tau \psi}c_{\mu \psi}f_{13}+c_{\mu \psi}^2f_{33}\right)\nn
    &+\frac{2\,\mathrm{Im}(c_{11})}{|c_{11}|^2}\left(c_{\tau s}c_{\tau \psi}f_{11}+\left(c_{\tau s}c_{\mu \psi}+c_{\mu s}c_{\tau \psi}\right)f_{13}+c_{\mu s}c_{\mu \psi}f_{33}\right)\nn
    &+\frac{1}{|c_{11}|^2}\left(c_{\mu s}\mathrm{Im}(c_{11}^\ast d_{11}))+c_{\mu\psi}\mathrm{Re}(d_{11})-c_{\tau s}\mathrm{Im}(c_{11}^\ast d_{21})-c_{\tau \psi}\mathrm{Re}(d_{21})\right)+\mathcal{O}(|\psi|^2)\,,\nn\nn
    Z_1&=(\partial_\epsilon p_0)^2f_{11}+f_{22}+(\partial_{\varrho}p_0)^2f_{33} +2f_{12} \partial_\epsilon p_0-2 \partial_\epsilon p_0\, \partial_\varrho p_0f_{13} +2\partial_\varrho p_0f_{23}+\mathcal{O}(|\psi|^2)\,,\nn
    Z_2&=Z_3=\frac{1}{2\,c_{11}}\left(\partial_\varrho p_0\,d_{11}+\partial_\epsilon p_0\,d_{21}-d_{22}\right)+i c_{\tau s}\left(\partial_\epsilon p_0\, f_{11}+f_{12}-\partial_\varrho p_0\, f_{13}\right)\nn&+\frac{1}{c_{11}}\left(c_{\tau\psi}\partial_\epsilon p_0\,f_{11}-c_{\mu\psi}\partial_\varrho p_0\, f_{33}+c_{\tau\psi}f_{12}-c_{\mu\psi}f_{23}+\left(c_{\mu\psi}\partial_\epsilon p_0-c_{\tau\psi}\partial_\varrho p_0\right)f_{13}\right)\nn&-i c_{\mu s}\left(\partial_\varrho p_0f_{33}+f_{23}-\partial_\epsilon p_0 f_{13}\right)+\mathcal{O}(|\psi|^2)\,,\nn
Z_\pi&=\beta^2\,\frac{e_{22}}{c_{11}^2}\,,\quad w_0=e_{11}\,,\nn
\sigma&=\frac{1}{(\epsilon_0+p_0)^2}\left(\beta \varrho_0^2\, r_{11}-2\varrho_0\beta(\epsilon_0+p_0)r_{12}+\beta (\epsilon_0+p_0)^2\, r_{22}\right)\,.
\end{align}
\endgroup

We note that the expression for $\Gamma_0$ includes terms up to order $\ves^2\sim|\psi|^2$, and the expressions for $Z_n,\, Z_1,\, Z_2,\,Z_\pi$ are $\mathcal{O}(\ves^0)\sim\mathcal{O}(|\psi|^0)\,$. After fixing the hydrodynamic frame and only taking into account the derivative expansion, we can also find the expressions for $\Gamma_0,\,Z_n,\, Z_1,\, Z_2$, for arbitrary $|\psi|^2$, in terms of the coefficients of $I_{EFT}$. We shall not present these results here, as they are quite lengthy. However, two important observations on these general, non-perturbative in $|\psi|^2$ expressions are in order. First, we have verified that the Onsager relations \eqref{eq:Recipr} are indeed satisfied even non-perturbatively. Second, we have checked that the inequalities \eqref{eq:Ineqs_2} (following from the positivity of entropy production) are indeed satisfied, upon employing the inequality constraints on the coefficients of $I_{EFT}$, which follow from the third unitarity condition \eqref{eq:unitarity}(see Appendix \ref{app:Ineq_unit}). 

As a final and more general comment regarding the Keldysh-Schwinger framework, we should emphasize that the equality and inequality-type constraints on transport coefficients are a direct consequence of unitarity and the dynamical KMS invariance of the Keldysh-Schwinger effective action \cite{Glorioso:2017fpd}. In contrast, in the case of standard hydrodynamics, these constraints are imposed by hand.

\section{Discussion}\label{sec:Disc}

In this paper, we constructed a relativistic effective theory for the nearly critical region of superfluids, up to next-to-leading order, in a specific perturbative scheme outlined in Section \ref{sec:Eff_th}. Compared to previous work \cite{KhalatnikovLebedev1978} in an appropriate non-relativistic limit, our theory predicts an extra complex coefficient $Z_\pi$ in the order parameter equation of motion. In a companion paper \cite{Donos:2025igh}, we have extracted the mean field theory limit of the same system from holography. Crucially, we confirmed that this extra term is necessary to capture the dynamics predicted by holographic theories. Given the microscopic nature of this calculation, it provides a valid testing ground for the completeness of effective theories.

In Section \ref{sec:Asymptotic}, we linearised the effective theory and discussed its asymptotic limits, for small and large momenta compared to the gap of the amplitude mode. In Subsection \ref{sec:Integr_out} we obtained the IR limit, where we have recovered superfluids at first order in the derivative expansion.  In Subsection \ref{sec:Normal_fluid} we examined the UV limit, and found normal fluid hydrodynamics, as expected.

Finally, in Section \ref{sec:KS_hydro} we presented an alternative construction of the theory in the framework of the Keldysh-Schwinger formalism. Together with the holographic derivation \cite{Donos:2025igh}, these constructions provide further justification for the effective theory that we propose. A particular point that provided invaluable input is the construction of the entropy current for which the Keldysh-Schwinger formalism offers an unambiguous recipe.

We studied different aspects of our theory in a mean field theory approximation. In Subsection \ref{sec:Integr_out}, after taking an appropriate IR limit of linearised fluctuations, we derived superfluid hydrodynamics with transport coefficients written in terms of our critical theory parameters. As we saw, the three bulk viscosities blow up in the critical region, exhibiting mean field theory behaviour. This was already noticed in the context of Model F and holography \cite{Donos:2022www,Donos:2022qao,Donos:2023ibv}. Moreover, we found that the diffusion constant $D_H$, appearing in the Higgs mode dispersion relation, exhibits different behaviour from the prediction of Model F, which decouples the order parameter sector from the normal fluid. An interesting future direction would be to study the effects of thermal fluctuations, within our Keldysh-Schwinger construction. Another possible next step would be to apply similar methods to study the critical dynamics of systems with spontaneously broken translations \cite{PhysRevD.99.086012}. To our knowledge, a complete construction of the effective theory has not been attempted in the past.

\section*{Acknowledgements}

AD and PK are supported by the Leverhulme Research Project Grant RPG-2023-058. AD is supported by STFC grant ST/T000708/1.

\appendix

\section{Non-redundancy of $Z_\pi$}\label{app:Elim_Z_pi}

In this appendix, we will argue that the coefficient $Z_\pi$ is non-redundant after using the leading order equations of motion and reshuffling the transport coefficients in $E_{\mathrm{diss}}$ \eqref{eq:corr_constit}.
More specifically, using the chain rule we find,
\begin{align}
    \label{eq:dt_F}\hat{D}_u\mathcal{F}_\psi=\left(-\partial_{|\psi|^2}\varrho\,\partial_u\mu-\partial_{|\psi|^2}s\,\partial_u T+\partial^2_{|\psi|^2}f\,\partial_u |\psi|^2\right)\psi+\partial_{|\psi|^2}f\, \hat{D}_u\psi-\frac{\hat{D}_u s_\psi}{2}\,,
\end{align}
where we omitted terms of order $\mathcal{O}(\ves^7)$ in the scaling region $\lambda\sim\ves^2$ and all terms appearing on both sides are of order $\mathcal{O}(\ves^5)$\footnote{Notice that $\partial^2_{|\psi|^2}f=\mathcal{O}(\ves^0)$, whereas $\partial_{|\psi|^2}f= \mathcal{O}(\ves^2)$.}. Using the leading equations of motion \eqref{eq:lead_cont} and \eqref{eq:lead_op}, we can trade all terms involving dynamical variables on the right hand side of \eqref{eq:dt_F} for terms involving $\nabla_\mu u^\mu$ and $\mathcal{F}_\psi$ only. Finally, substituting this expression for $\hat{D}_u\mathcal{F}_\psi$ back in \eqref{eq:corr_constit}, we find the equivalent expression,
\begin{align}\label{eq:Ediss_alt}
    E_{diss}=&-2\,\overline{\Gamma}'_0\,\mathcal{F}_\psi-Z_n'\, \psi^2\, \mathcal{F}_\psi^\ast+Z_2'\,\psi\, \nabla_\mu u^\mu+Z_\pi\,\frac{\hat{D}_u s_\psi}{2}\,.
\end{align}
The primed coefficients are related to the original ones according to,
\begin{align}
    \label{eq:primed}\Gamma_0'&=\Gamma_0+\frac{|\psi|^2}{2}\overline{Z}_\pi\,  \frac{\kappa_a}{\kappa}-\Gamma_0\,\overline{Z}_\pi\,\partial_{|\psi|^2}f,\quad
    Z_n'=Z_n+Z_\pi\,\frac{\kappa_a}{\kappa}\,,\quad
    Z_2'=Z_2-Z_\pi\, \frac{\kappa_b}{\kappa}\,,
\end{align}
with,
\begin{align}
   \label{eq:kappas}\kappa_a&=2\,\Gamma_0\left(\partial_{|\psi|^2}\epsilon\left(\partial_\mu\varrho\, \partial_{|\psi|^2}s-\partial_T\varrho\,\partial_{|\psi|^2}\varrho\right)-\partial_\mu\epsilon\left(\partial_{|\psi|^2}\varrho\,\partial_{|\psi|^2}s+\partial_T\varrho\, \partial^2_{|\psi|^2}f\right)\right)\nn
   &+2\,\Gamma_0\,\partial_T\epsilon\left(\partial_\mu\varrho\,\partial^2_{|\psi|^2}f+(\partial_{|\psi|^2}\varrho)^2\right)+iq_e\partial_{|\psi|^2}s\left(\partial_\mu \epsilon-\mu\, \partial_\mu\varrho\right)+iq_e\partial_{|\psi|^2}\varrho\left(\mu\,\partial_T\varrho-\partial_T\epsilon\right)\,,\nn
   \kappa_b&=(\epsilon+p)\left(\partial_{|\psi|^2}\varrho\,\partial_T\varrho-\partial_{|\psi|^2}s\,\partial_\mu\varrho\right)+\varrho\left(\partial_{|\psi|^2}s\,\partial_\mu\epsilon-\partial_{|\psi|^2}\varrho\,\partial_T\epsilon\right)\,,\nn
   \kappa&=\partial_\mu\epsilon\,\partial_T\varrho-\partial_T\epsilon\,\partial_\mu\varrho\,.
\end{align}

The expressions for the new coefficients are not particularly important. What is crucial, though, is that $\mathrm{Im}(Z'_n)\neq0$ and $Z_2'\neq Z_2=Z_3$ which would naively appear to be in contrast with the Onsager reciprocity constraints \eqref{eq:Recipr} that we discussed in the main text. However, the extra term involving $Z_\pi$ and the time derivative of the source $s_\psi$ in \eqref{eq:Ediss_alt}, guarantees that Onsager reciprocity is restored. If such a term had been included in the analysis of \cite{KhalatnikovLebedev1978}, the authors would have concluded that the constraint \eqref{eq:Recipr}, which they also have, could have been replaced by,
\begin{align}
\mathrm{Im}(Z'_n)=\frac{\mathrm{Im}(\kappa_a\,Z_\pi)}{\kappa}\,,\quad Z_2'=Z_3-Z_\pi\frac{\kappa_b}{\kappa}\,.
\end{align}
The coefficient $\kappa_a$ can be expressed in terms of $\Gamma_0'$ using \eqref{eq:primed} and \eqref{eq:kappas}.

\section{Definitions of susceptibilities}\label{app:susc_def}

In this appendix, we define various susceptibilities that we found useful in our construction. In particular, we are interested in thermal equilibrium states, which are spatially homogeneous and isotropic, without gradients and external sources. Given the free energy of the system $f$ for fixed temperature $T$, chemical potential $\mu$ and condensate modulus $|\psi|$, we define its conjugate variable $\pi$ as in \eqref{eq:pi_def}, 
\begin{align}\label{eq:pi_app}
  \pi=\left(\frac{\partial f}{\partial |\psi|}\right)_{\mu,T}\,.  
\end{align}
When the system is strictly at thermodynamic equilibrium, the conjugate variable $\pi$ is identically zero. In near equilibrium, we imagine a manifold of partial equilibrium states  with coordinates $T,\mu, |\psi|$, in which $\pi$ can be a non-zero but parametrically small quantity.

Notice that the mapping $|\psi|\to\pi(|\psi|)$, for fixed $\mu$ and $T$, is not injective for our systems in their broken phase. Solving \eqref{eq:pi_app} for $|\psi|$ must give us at least two solutions, $|\psi_{\ast}(\mu,T,\pi)|$ and $|\psi_{\#}(\mu,T,\pi)|$, corresponding to the spontaneously broken phase and the normal phase of the system, respectively,
\begin{align}\label{eq:psi_branches}
    |\psi_{\ast}(\mu,T,\pi=0)|=\rho_v(\mu,T),\quad|\psi_{\#}(\mu,T,\pi=0)|=0\,.
\end{align}
Moreover, after a simple Legendre transformation of the free energy $f$ we can also define the broken and normal phase partial equilibrium charge and entropy densities, as functions of $\pi$,
\begin{align}
    \varrho_{\ast}(\mu,T,\pi)=\varrho(\mu,T,|\psi_{\ast}(\mu,T,\pi)|)\,,\quad \varrho_{\#}(\mu,T,\pi)=\varrho(\mu,T,|\psi_{\#}(\mu,T,\pi)|)\,,\nn
    s_{\ast}(\mu,T,\pi)=s(\mu,T,|\psi_{\ast}(\mu,T,\pi)|)\,,\quad s_{\#}(\mu,T,\pi)=\varrho(\mu,T,|\psi_{\#}(\mu,T,\pi)|)\,.
\end{align}
The two different branches are defined by the two distinct solutions \eqref{eq:psi_branches} we can write for $|\psi|$ when inverting \eqref{eq:pi_app}.

Given the above thermodynamic quantities, the first set of susceptibilities we would like to define are relevant to the amplitude of the order parameter, 
\begin{align}\label{eq:defs_v1}
    \nu_{\mu\rho}&=\left(\frac{\partial \rho_v(\mu,T)}{\partial \mu}\right)_T,\quad\nu_{T\rho}=\left(\frac{\partial \rho_v(\mu,T)}{\partial T}\right)_\mu\,,\nn\nu_{\rho\rho}&=\left(\frac{\partial |\psi_\ast(\mu,T,\pi)|}{\partial \pi}\right)_{\mu,T, \pi=0},\quad
    \nu^\#_{\rho\rho}=\left(\frac{\partial |\psi_\#(\mu,T,\pi)|}{\partial \pi}\right)_{\mu,T,\pi=0}.
\end{align}
In order to define $\nu_{\rho\rho}$ and $\nu^\#_{\rho\rho}$ we considered the system to be out of equilibrium with $\pi\neq 0$, . However, these quantities are actually thermodynamic since this would be equivalent to the susceptibilities that we would define by considering external sources for the order parameter. Finally, we can also define the more standard susceptibilities,
\begin{align}
    \chi&=\left(\frac{\partial \varrho_\ast}{\partial \mu}\right)_{T,\,\pi=0}=\left(\frac{\partial \varrho(\mu,T,\rho_v(\mu,T))}{\partial\mu}\right)_T\,,\quad\chi^\#=\left(\frac{\partial \varrho_\#}{\partial \mu}\right)_{T,\,\pi=0}=\left(\frac{\partial \varrho(\mu,T,|\psi|=0)}{\partial\mu}\right)_T\,,\nn
\xi&=\left(\frac{\partial \varrho_\ast}{\partial T}\right)_{\mu,\,\pi=0}=\left(\frac{\partial \varrho(\mu,T,\rho_v(\mu,T))}{\partial T}\right)_\mu\,,\quad\xi^\#=\left(\frac{\partial \varrho_\#}{\partial T}\right)_{\mu,\,\pi=0}=\left(\frac{\partial \varrho(\mu,T,|\psi|=0)}{\partial T}\right)_\mu\,,\nn
\frac{c_\mu}{T}&=\left(\frac{\partial s_\ast}{\partial T}\right)_{\mu,\,\pi=0}=\left(\frac{\partial s(\mu,T,\rho_v(\mu,T))}{\partial T}\right)_\mu\,,\quad\frac{c_\mu^\#}{T}=\left(\frac{\partial s_\#}{\partial T}\right)_{\mu,\,\pi=0}=\left(\frac{\partial s(\mu,T,|\psi|=0)}{\partial T}\right)_\mu\,,
\end{align}
which only involve derivatives with respect to the temperature and chemical potential.

After our formal definitions, we also comment on a couple of identities that we found useful. Notice that $\varrho,s$ appearing in \ref{sec:Eff_linear} are essentially $\varrho_{\ast}(\mu,T,\pi),\,s_\ast(\mu, T, \pi)$. Also, using the chain rule we can write,
\begin{align}
   & \left(\frac{\partial \varrho_{\ast}(\mu,T,\pi)}{\partial \pi}\right)_{\mu,T}=\left(\frac{\partial \varrho(\mu,T,|\psi|=|\psi_{\ast}|)}{\partial |\psi|}\right)_{\mu,T}\left(\frac{\partial |\psi_{\ast}|}{\partial \pi}\right)_{\mu,T}=\nn&=-\left(\frac{\partial \pi(\mu,T,|\psi|)}{\partial \mu}\right)_{T, |\psi|=|\psi_\ast|}\left(\frac{\partial |\psi_{\ast}|}{\partial \pi}\right)_{\mu,T}=\left(\frac{\partial |\psi_{\ast}|}{\partial \mu}\right)_{\pi,T}\,,
\end{align}
and after evaluating this at $\pi=0$ we obtain,
\begin{align}\label{eq:nu_murhoalt}
     \left(\frac{\partial \varrho_{\ast}(\mu,T,\pi=0)}{\partial \pi}\right)_{\mu,T}=\left(\frac{\partial |\psi_{\ast}|}{\partial \mu}\right)_{\pi=0,T}=\left(\frac{\partial \rho_v}{\partial \mu}\right)_{T}=\nu_{\mu\rho}.
\end{align}
By following similar steps, we can also write,
\begin{align}\label{eq:nu_Trhoalt}
    \left(\frac{\partial s_{\ast}(\mu,T,\pi=0)}{\partial \pi}\right)_{\mu,T}=\nu_{T\rho}\,.
\end{align}

\section{Relations close to $T_c$}\label{app:mean_exp}

 In this appendix, we discuss the behaviour of susceptibilities and other thermal quantities close to the critical point, based on a mean field theory approximation \cite{Goodstein2002}. Before discussing this approximation, we assume that the Landau free energy $f$ is an analytic function of temperature $T$, the chemical potential $\mu$ and the order parameter's modulus, $|\psi|^2$. This implies that close to the critical point it admits an expansion in powers of $|\psi|^2$,
\begin{equation}
f(\mu,T,|\psi|^2)=f_0+\beta\,|\psi|^2+\gamma\,|\psi|^4+\cdots\,,
\end{equation}
with $f_0,\beta,\gamma$ functions of $\mu,T$ only and $\beta\leq0,\,\gamma>0$. Without a condensate, we would only be left with $f_0$, which is naturally interpreted as the free energy of the normal phase.

Close to the critical point, in order to describe a second order phase transition, one further assumes that for $\mu(\ves)$, $T(\ves)$ as in \eqref{eq:mu_T}, the coefficients behave as
\begin{align}
    \beta&=\beta_{0,\mu}\,(\mu-\mu_c)+\beta_{0,T}\,(T-T_c)+\mathcal{O}(\ves^4)\,,\nn
    \gamma&=\gamma_0+\mathcal{O}(\ves^2)\,,
\end{align}
and $|\psi|\approx\mathcal{O}(\ves)$. From the definitions \eqref{eq:therm_def}, \eqref{eq:pi_app}, we immediately find,
\begin{align}\label{eq:mean_rels}
    \varrho(\mu,T,|\psi|)&=\varrho_0(\mu,T)-\beta_{0,\mu}|\psi|^2+\mathcal{O}(\ves^4)\,,\nn
    s(\mu,T,|\psi|)&=s_0(\mu,T)-\beta_{0,T}|\psi|^2+\mathcal{O}(\ves^4)\,,\nn
   \pi(\mu,T, |\psi|)&=2|\psi|\left(\beta_{0,\mu}(\mu-\mu_c)+\beta_{0,T}(T-T_c)+2\gamma_0 |\psi|^2\right)+\mathcal{O}(\ves^5)\,,
\end{align}
with $\varrho_0=-\left(\frac{\partial f_0}{\partial\mu}\right)_T\,,s_0=-\left(\frac{\partial f_0}{\partial T}\right)_\mu$. Notice that $|\psi_\#(\mu,T,\pi=0)|=0$ and   $|\psi_{\ast}(\mu,T,\pi=0)|=\rho_v=\sqrt{-\frac{\beta_{0,\mu}(\mu-\mu_c)+\beta_{0,T}(T-T_c)}{2\gamma_0}}+\mathcal{O}(\ves^3)\,.$

Taking the derivative of the third relation in \eqref{eq:mean_rels} with respect to $\pi$, according to the definitions of the previous subsection, we find that
\begin{align}
    \nu_{\rho\rho}&=-\frac{1}{4 \left(\beta_{0,\mu}(\mu-\mu_c)+\beta_{0,T}(T-T_c)\right)}+\mathcal{O}(\ves)\,,\nn
    \nu_{\rho\rho}^\#&=\frac{1}{2 \left(\beta_{0,\mu}(\mu-\mu_c)+\beta_{0,T}(T-T_c)\right)}+\mathcal{O}(\ves)=-2\, \nu_{\rho\rho}+\mathcal{O}(\ves)\,.
\end{align}
Also, using either the definitions \eqref{eq:defs_v1} or \eqref{eq:nu_murhoalt}, \eqref{eq:nu_Trhoalt} we can show that
\begin{align}
\nu_{\mu\rho}=-2\,\rho_v\,\nu_{\rho\rho}\,\beta_{0,\mu}+\mathcal{O}(\ves)\,,\quad \nu_{T\rho}=-2\,\rho_v\,\nu_{\rho\rho}\,\beta_{0,T}+\mathcal{O}(\ves)\,.
\end{align}
Taking the derivative of the first of \eqref{eq:mean_rels}, evaluated at $|\psi|=\rho_v$, we get,
\begin{align}\label{eq:NP_1}
    \left(\frac{\partial \varrho(\mu,T,\rho_v(\mu,T))}{\partial\mu}\right)_T&=\left(\frac{\partial \varrho_0(\mu,T)}{\partial\mu}\right)_T-2\beta_{0,\mu}\rho_v \left(\frac{\partial \rho_v}{\partial \mu}\right)_T+\mathcal{O}(\ves^2)\Rightarrow\nn
    \quad\chi&=\chi^\#+\frac{\nu_{\mu\rho}^2}{\nu_{\rho\rho}}+\mathcal{O}(\ves^2)\,.
\end{align}
Similarly, it follows that,
\begin{align}\label{eq:NP_2}
    \frac{c_\mu}{T}=\frac{c_\mu^\#}{T}+\frac{\nu_{T\rho}^2}{\nu_{\rho\rho}}+\mathcal{O}(\ves^2)\,,\quad
     \xi=\xi^\#+\frac{\nu_{T\rho}\,\nu_{\mu\rho}}{\nu_{\rho\rho}}+\mathcal{O}(\ves^2)\,.
\end{align}

\section{Normal fluid corrections}\label{app:norm_fluid_tensors}\label{app:nf_corrections}
In this appendix we give explicit expressions for the corrections that appear in the Lagrangian term $\mathcal{L}_{n=2}$ of equation \eqref{eq:L_2} and which have appeared in the literature before. In particular, for the quantity $W_0^{\mu\nu,M N}$ that appears in \eqref{eq:L_2}, we can write,
\begin{align}
W_0^{\mu\alpha,\nu\beta}&= s_{11}u^\mu u^\nu u^\alpha u^\beta+ s_{22}P^{\mu\nu}P^{\alpha\beta} - s_{12}(u^{\mu} u^{\nu}P^{\alpha\beta}+u^{\alpha} u^{\beta}P^{\mu\nu}) \nn
&\ \ + 2 r_{11}\left(u^\mu u^{(\alpha}P^{\beta)\nu}+u^\nu u^{(\alpha}P^{\beta)\mu}\right)+ 4 r
\left(P^{\alpha(\mu}P^{\nu)\beta} - \frac{1}{d-1}P^{\mu\nu}P^{\alpha\beta}  \right)\nn
W_0^{\mu\alpha,\nu d}&=-  s_{13}u^\mu u^\nu u^\alpha+   s_{23}P^{\mu\nu} u^\alpha+ 2 r_{12}
u^{(\mu}P^{\nu)\alpha} , \quad  W_0^{\mu\alpha,d \nu} =  W_0^{\alpha \mu,\nu d} \nn
\label{W02}W_0^{\mu\nu,dd}&= s_{33}u^\mu u^\nu+ r_{22}P^{\mu\nu} \ .
\end{align}
We have directly imported this structure from \cite{Glorioso:2017fpd} with the only difference that the coefficients $s_{ab}, r$ etc., can also depend on the order parameter modulus $|\hat{\psi}|^2$ besides the temperature and chemical potential.

In the main text, we decomposed the first derivative corrections to the stress tensor and the $U(1)$ current to a normal fluid and an order parameter contribution. For the normal fluid corrections we can write \cite{Glorioso:2017fpd},
\begin{align}\label{1gens}
T_{1,n}^{\mu \nu} &= h_{\epsilon}\, u^\mu u^\nu + h_{p}\, P^{\mu \nu} +
2\,u^{(\mu} q_1^{\nu)} -\eta\, \sigma^{\mu \nu}\,,\nn
J_{1,n}^\mu &= h_n \,u^\mu + j_1^\mu\,.
\end{align}
In the expressions above we have defined,
\begin{align}
  h_\epsilon &=
f_{11}\, \partial_u \tau +  f_{12}\, \nabla_\mu u^\mu +  \frac{f_{13}}{\beta}\, \partial_u \hat{\mu}\,, \nn
h_p &=  f_{21}\, \partial_u \tau - f_{22}\, \nabla_\mu u^\mu +  \frac{f_{23}}{\beta}\, \partial_u \hat{\mu}\,, \nn
h_n  &= f_{31} \partial_u \tau + f_{32}\, \nabla_\mu u^\mu -   \frac{f_{33}}{\beta} \, \partial_u \hat{\mu}\,,  \nn
 j_1^\mu &= \lambda_{21} \partial_u u^\mu
 -\lambda_2  \left( P^{\mu \nu} \partial_\nu  \mu + u_\lambda  F^{\lambda \mu} \right)+ \lambda_7 P^{\mu \nu} \partial_\nu  \tau +\lambda_{8} P^{\mu \nu} \partial_\nu  \mu\,, \nn
 q_1^\mu &= - \lambda_{1}  \partial_u u^\mu  +\lambda_{12}  \left( P^{\mu \nu} \partial_\nu  \mu + u_\lambda  F^{\lambda \mu} \right)+ \lambda_5 P^{\mu \nu} \partial_\nu  \tau +\lambda_{6} P^{\mu \nu} \partial_\nu  \mu\,.
\end{align}
Once again, all the parameters that appear in these expressions can be in principle depend on $T$, $\mu$ and $|\hat{\psi}|^2$. For completeness, we also list the KMS constraints that these have to satisfy,
\begin{align}
    \label{eq:NOrmal_rel}&\lambda_5=\lambda_1+\mu\lambda_{12}\,,\quad \lambda_7=-\lambda_{21}-\mu\lambda_2,\,\quad\lambda_6=\lambda_8=0\,,\nn
    &\lambda_{12}=\lambda_{21},\,\quad f_{31}=-f_{13}\,,\quad f_{32}=f_{23}\,,\quad f_{21}=-f_{12}\,,\nn
    &r=\frac{\eta}{2} T\,,\quad r_{11}=\lambda_1\,T\,,\quad r_{12}=-\lambda_{12}\,T,\quad r_{22}=\lambda_2\,T\,,\nn
    &s_{11}=f_{11}\,T,\,\quad s_{12}=f_{12}\,T\,,\quad s_{13}=f_{13}\,T\,,\nn &s_{22}=f_{22}\,T\,,\quad s_{23}=-f_{23}\,T\,,\quad s_{33}=f_{33}\,T\,.
\end{align}

\section{Inequalities from unitarity}\label{app:Ineq_unit}

The third of the unitarity conditions \eqref{eq:unitarity} yields several inequality relations among the coefficients of $\mathcal{L}_{EFT}$. For $n=1$ we have $\mathrm{Im}\left(\mathcal{L}_{[1]}\right)=0$ and the constraint is trivially satisfied. For $n=2$, taking into account \eqref{eq:KMS_2}, we have
\begin{align}\label{eq:im_l2}\mathrm{Im}\left(\mathcal{L}_{[2]}\right)&=\frac{1}{4}W_0^{\mu\nu,M N}G_{a\mu M}G_{a\nu N} +2\,\mathrm{Im}\left(\lambda_0^{\mu M}\right)G_{a \mu M}\, \mathrm{Re}\left[\hat{\psi}_a\,\hat{\psi}^\ast\right]\nn&+2\,\mathrm{Im}\left(\kappa_{1}\right)\mathrm{Re}\left[\left(\hat{\psi}_a \hat{\psi}^\ast\right)^2\right]+ \frac{\kappa_{0}}{|\hat{\psi}|^2}|\hat{\psi}_a\hat{\psi}^\ast|^2\nn &\equiv \frac{1}{\beta}Q_{S,1}+\frac{1}{\beta}Q_{S,2}+Q_V+Q_T,\,
\end{align}
where the $Q$'s are the following quadratic forms
\begin{align}
Q_{S,1}&=f_{11}\,s_a^2+f_{22}\,s_b^2+f_{33}\,s_c^2+(-2\mathrm{Re}(c_{11})-2c_{15}|\hat{\psi}|^2)s_d^2-2f_{12}\,s_a\,s_b-2f_{13}\,s_a\,s_c-2f_{23}\,s_b\,s_c\nn&-2|\hat{\psi|}(\mathrm{Re}(d_{21})s_a+\mathrm{Re}(d_{22})s_b+\mathrm{Re}(d_{11})s_c)s_d\,,\nn
    Q_{S,2}&=(-2\mathrm{Re}(c_{11})+2c_{15}|\hat{\psi}|^2)s_e^2\,,
    \end{align}
    \begin{align}
    Q_V&=r_{11}V^\mu_A V_{A\mu}+r_{22}V^\mu_B V_{B\mu}+2r_{12}V^\mu_AV_{B\mu}\,,\nn
    Q_T&=r\,t^{\mu\nu}_A t_{A\mu\nu}\,,
\end{align}
with,
\begin{align}
s_a&=\frac{1}{2}u^\mu u^\nu G_{a\mu\nu}\,,s_b=\frac{1}{2}P^{\mu\nu}G_{a\mu\nu},\,s_c=u^\mu C_{a\mu}\,,
\nn s_d&=\frac{1}{|\hat{\psi}|}\mathrm{Re}(\hat{\psi}_a\hat{\psi}^\ast)\,,\quad s_e=\frac{1}{|\hat{\psi}|}\mathrm{Im}(\hat{\psi}_a\hat{\psi}^\ast)\,,\nn
V^\mu_A&=u^\nu P^{\mu\rho}G_{a\nu\rho}\,,\quad V^\mu_B=P^{\mu\nu}C_{a\nu},\nn
    t^{\nu\sigma}_A&=(P^{\mu\nu}P^{\rho\sigma}-\frac{1}{d-1}P^{\mu\rho}P^{\nu\sigma})G_{a\mu\rho}\,.
\end{align}
The four quadratic forms must all be separately semipositive definite for $\mathrm{Im}(\mathcal{L}_{[2]})\geq0$ to hold. The form of $Q_V$ and $Q_T$ is exactly the same as in the case of normal fluids \cite{Glorioso:2017fpd} and demanding their semipositivity, given the identification \eqref{eq:Rel_hydro_Sk}and \eqref{eq:NOrmal_rel}, immediately leads to $\sigma\geq0\,,\eta\geq0\,$. Requiring the semipositivity of $Q_{S,1},\,Q_{S,2}$ leads to 16, in total, inequalities for their coefficients\footnote{There is one condition from $Q_{S,2}$ and 15 conditions for the coefficients of $Q_{S,1}$, equal to the number of principal minors of the corresponding 4 by 4 matrix.}. Using  these inequalities and the (non-perturbative in $\ves$) expressions for the effective theory coefficients in terms of the coefficients of $I_{EFT}$, it is straightforward to check that the inequalities \eqref{eq:Ineqs_2} and \eqref{eq:Ineqs_1}, for $Z_1$, are obeyed. 

\newpage
\bibliographystyle{utphys}
\bibliography{refsthesis}{}
\end{document}